\documentclass[sigconf]{acmart}

\AtBeginDocument{%
  \providecommand\BibTeX{{%
    \normalfont B\kern-0.5em{\scshape i\kern-0.25em b}\kern-0.8em\TeX}}}

\copyrightyear{2023} 
\acmYear{2023} 
\setcopyright{rightsretained} 
\acmConference[MobiHoc '23]{The Twenty-fourth International Symposium on Theory, Algorithmic Foundations, and Protocol Design for Mobile Networks and Mobile Computing}{October 23--26, 2023}{Washington, DC, USA}
\acmBooktitle{The Twenty-fourth International Symposium on Theory, Algorithmic Foundations, and Protocol Design for Mobile Networks and Mobile Computing (MobiHoc '23), October 23--26, 2023, Washington, DC, USA}
\acmDOI{10.1145/3565287.3610258}
\acmISBN{978-1-4503-9926-5/23/10}

\usepackage{graphicx}
\usepackage{booktabs} 
\usepackage{amsmath}
\usepackage{amsfonts}

\usepackage[flushleft]{threeparttable}

\usepackage{tabularx}

\usepackage{epsf}
\usepackage{fancyhdr}
\usepackage{graphics}
\usepackage{graphicx}
\usepackage{psfrag}
\usepackage{microtype}

\usepackage[linesnumbered,ruled]{algorithm2e}

\usepackage{mathtools}

\usepackage{color}
\usepackage{amsthm}
\usepackage{amsfonts}
\usepackage{amsmath}
\usepackage{bbm}
\usepackage{caption,subcaption}
\usepackage{sidecap}
\sidecaptionvpos{figure}{c}

\usepackage{url}
\usepackage{hyperref}

\long\def\comment#1{}
\comment{
\setlength{\topmargin}{0 in}
\setlength{\textwidth}{5.5 in}
\setlength{\textheight}{8 in}
\setlength{\oddsidemargin}{0.5 in}
}

\theoremstyle{plain}

\newtheorem{lemma}{Lemma}
\newtheorem{theorem}{Theorem}

\newtheorem{definition}{Definition}

\newtheorem{assumption}{Assumption}

\newtheorem{corollary}{Corollary}

\def\EE{\mathbb{E}}

\def\WQQ{\widetilde{\mathbf{Q}}}

\def\QQ{\mathbf{Q}}

\def\WW{\mathbf{W}}

\def\Rr{\mathcal{R}}

\def\Mm{\mathcal{M}}

\def\Cc{\mathcal{C}}

\def\Aa{\mathcal{A}}

\def\thetav{\boldsymbol{\theta}}

\def\aA{\mathbf{a}}
\def\bB{\mathbf{b}}

\def\tq{\widetilde{Q}}

\def\xX{\mathbf{x}}

\def\pP{\mathbf{p}}

\def\pP{\mathbf{p}}

\def\Lmb{\mathbf{\Lambda}}

\def\lmb{\boldsymbol{\lambda}}

\def\wlmb{\tilde{\boldsymbol{\lambda}}}

\def\muu{\boldsymbol{\mu}}

\graphicspath{{./image/}}

\DeclareMathOperator*{\argmaxE}{\arg\!\max}

\newcommand{\argmax}{\mathop{\arg \max}}

\newcommand{\ba}{\begin{array}}
\newcommand{\ea}{\end{array}}

\makeatletter
\DeclareRobustCommand{\Div}{\mathrel{\Div@}}
\DeclareRobustCommand{\nDiv}{\mathrel{\nDiv@}}

\newcommand{\nDiv@}{\centernot\Div@}
\makeatother

\usepackage{centernot}

\begin{document}

\title{Learning to Schedule in Non-Stationary Wireless Networks With Unknown Statistics}

\author{Quang Minh Nguyen}
\email{nmquang@mit.edu}
\affiliation{%
  \institution{Massachusetts Institute of Technology}
   \country{USA}}

\author{Eytan Modiano}
\email{modiano@mit.edu}
\affiliation{%
  \institution{Massachusetts Institute of Technology}
  \country{USA}}


\begin{abstract}
The emergence of  large-scale wireless networks with partially-observable and time-varying dynamics has imposed new  challenges on the design of optimal control policies. This paper studies efficient scheduling algorithms for wireless networks subject to generalized interference constraint, where mean arrival  and mean service rates are unknown and  non-stationary. This model exemplifies realistic edge devices' characteristics of wireless communication in modern networks. 
We propose a novel  algorithm termed MW-UCB for generalized wireless network scheduling, which is based on the Max-Weight policy and leverages the Sliding-Window Upper-Confidence Bound to learn the channels' statistics under non-stationarity. MW-UCB  is provably throughput-optimal under mild assumptions on the  variability of mean service rates.  Specifically, as long as the total variation in mean service rates over any time period grows sub-linearly in  time,  we show that MW-UCB can achieve the stability region arbitrarily close to the stability region of the class of policies with full knowledge of the channel statistics. Extensive simulations  validate our theoretical results and demonstrate the favorable performance of MW-UCB.


 
\end{abstract}

\ccsdesc[500]{Networks~Network performance modeling}
\ccsdesc[500]{Networks~Network performance analysis}
\ccsdesc[500]{Networks~Mobile ad hoc networks}
\ccsdesc[500]{Networks~Network control algorithms}

\keywords{Optimal Control, Scheduling, Wireless Network, Machine Learning, Partial Observability, Non-stationarity}



\maketitle

\section{Introduction}

Wireless networks  are increasingly large-scale and complex in response to the surge in edge-based Internet of Things (IoT) architecture \cite{edgeIOT, fedxgboost}, mobile communication \cite{wirelessSDN1}  and wireless paradigm  \cite{sdn_wireless}. 
One fundamental challenge in the transition to large-scale networks is that minor inefficiencies can accumulate and severely limit performance \cite{scala_wireless1}. 
Consequently, the advance of modern infrastructures toward  massive scale has  led to the  re-design of operational management for various network tasks,  such as  
 traffic engineering \cite{large_scale_te1}, load-balancing \cite{large_scale_load_balance1}, utility maximization \cite{dist_util_sdn}, and link scheduling  \cite{5gscheduling1, STAHLBUHK2019131}.
 In this work, we focus on designing scheduling algorithms that are theoretically efficient and meet the stringent requirements of emerging large-scale wireless networks.

 Efficient scheduling of transmissions is essential for wireless devices  to share the common spectrum while achieving high throughput.
Despite its established throughput-optimality for a variety of  classical stochastic network models,  the celebrated Max-Weight scheduling policy \cite{neelybook1, BP1} requires the full knowledge of the channel statistics, which are often  unknown a priori  \cite{STAHLBUHK2019131, bipartite_queueing2} and thus hinder its direct adoption. 
 First, due to the delay incurred by the accumulation of global network state information in emerging large-scale systems and multi-path fading, the instantaneous service capacities of wireless links and the packet arrivals to nodes are usually unavailable at the time of making scheduling decisions and can only be observed from channel feedback. We refer to this peculiar characteristic of large-scale networked systems as \emph{partial observability}. Second, the mobility  of edge devices \cite{edge_nonstationary1, volatile_edge1} and  unreliable nature of wireless communication \cite{unreliable_strong_consistency1} impose  \emph{non-stationary dynamics}, whereby both the mean packet arrivals and mean service rates may vary over time, and are \emph{unknown} in advance to the network operator.
When the channel is not instantaneously observable,   it is well-known that an optimal  policy is to leverage  the mean service rates in making Max-Weight  scheduling decisions \cite{neelybook1, BP1}; however, in our setting, those statistics are \emph{unknown}, \emph{non-stationary} and must be learned.
 In this paper, we aim to develop throughput-optimal scheduling algorithms  under  the  requirements of  \emph{partial observability}, \emph{non-stationary dynamics} and \emph{unknown statistics}.

 





A main challenge in the design of non-stationary network control algorithm under partially-observed and unknown statistics is that the analytical characterization of the capacity region for stationary network setting \cite{neelybook1} no longer holds under non-stationarity due to the potential non-existence of  steady state or well-defined long-term averages \cite{network_nonstationary1}. 
Previous works either consider simplified models \cite{STAHLBUHK2019131, adversarial_control1}, or only achieves a  constrained stability region for   bipartite queueing system \cite{bipartite_queueing2}. In particular, under \emph{partial observability} and \emph{unknown statistics}, \cite{STAHLBUHK2019131}  designed a throughput-optimal joint learning and scheduling policy for stationary network control. While establishing the effectiveness of the Max-Weight policy even for non-stationary network control, \cite{adversarial_control1} assumes the availability of instantaneous nodes' packet arrivals and links' service capacities to the controller for making decision. Closest to our work is \cite{bipartite_queueing2}, which proposes a stabilizing algorithm for bipartite queueing system that supports arrival rates within a stability region constrained by window-based (non-stationary) dynamics.



In this paper, we propose a new notion of stability for non-stationary network control, and  a novel joint learning and scheduling algorithm that achieves a stability region arbitrarily close to the true stability  region.
Our contributions can be summarized as follows:
\begin{itemize}
    \item We present a new class of approximate stability regions that is parameterized by a quantity capturing the closeness to  the true stability region. Based on this notion of approximate stability region, we propose a new notion of throughput-optimality for non-stationary network control and, as a special case, prove its equivalence to the conventional notion of stability in the simplified setting of stationary network. 
    \item We propose Max-Weight scheduling augmented by  Sliding-Window Upper-Confidence Bound, hence termed MW-UCB, as a novel algorithm  for non-stationary network control, subject to generalized wireless interference constraints, with partial observability and unknown statistics. Under  mild assumptions on the system learnability,  we  establish the throughput-optimality of MW-UCB and  its strong stability within the window-based region previously considered in the literature \cite{bipartite_queueing2}.
    \item We empirically validate our theoretical results and demonstrate that MW-UCB achieves the same stability region as that of the idealized Max-Weight policy with full knowledge of network statistics.  
\end{itemize}
The rest of the paper is organized as follows. We present our system model and problem formulation in Section \ref{sec:sys_model}. In Section \ref{sec:notion_stability}, we present our new notion of throughput-optimality for non-stationary networks. In Section \ref{sec:main_algo}, we propose the throughput-optimal MW-UCB algorithm and establish its stability results. We conduct numerical simulations to empirically validate the throughput-optimality  of  MW-UCB and demonstrate its favorable performance in  Section \ref{sec:exp}, and conclude the paper in Section \ref{sec:conclusion}.

\section{Preliminaries and Problem Formulation}
\label{sec:sys_model}

\subsection{Network Model}

A wireless network with arbitrary topology is represented by a directed graph $\mathcal{G}(V, E)$, where $V=\{1,2,...,n\}$ is the set of nodes and $E = \{ (i, j) : i, j \in V \}$ is the set of directed point-to-point links.  
Time is slotted.
For simplicity of technical exposition, we consider  single-hop traffic\footnote{The results of the paper naturally generalize to multi-hop setting by incorporating the Back-pressure mechanism \cite{BP1}.}. For any  $e = (i, j) \in E$, we denote by $a_e(t)$ the number of packets arriving at node $i$ at time slot $t$  to be transmitted to neighbouring node $j$. We consider $\{a_e(t)\}_{t\geq0}$ to be independent with potentially time-varying means $\lambda_e(t) = \EE[a_e(t)]$, and are bounded by a finite number, i.e. $a_e(t) \leq A_{max}$ for all $e\in E$ and $t$.
\begin{figure}[htbp]
\centerline{\includegraphics[scale=0.4]{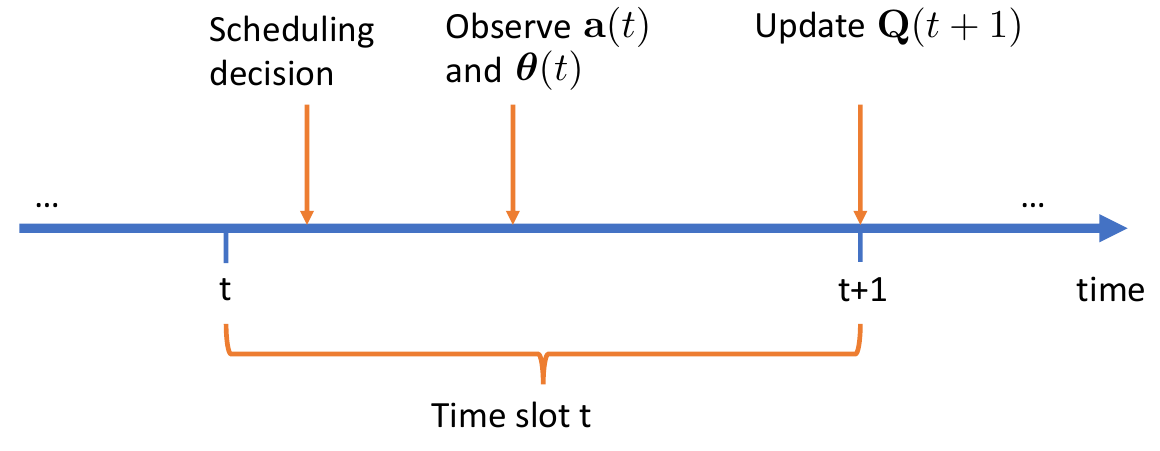}}
\caption{Sequence of events in one time slot.}
\label{fig: event_sequence}
\end{figure}

We assume a general  wireless interference model.  
Denote by  $\Mm$ the set of all admissible link activations and, at time slot $t$,  by  $\xX(t) = \{ x_e(t)\}_{e\in E} \in \Mm$  the  scheduling decision of whether to activate link $e \in E$: 
\begin{align*}
x_e(t) = 
    \begin{cases} 1  \text{, if $e$ is activated at time $t$} & \\
    0  \text{, if $e$ is not activated at time $t$} \end{cases}. 
\end{align*}
We impose no structural restriction on the set $\Mm$, thereby capturing a wide range of practical wireless models including  primary interference \cite{interference1}, k-hop interference \cite{k_hop}, and protocol interference \cite{protocol_interference}. Let $\theta_e(t)$ be the service capacity of link $e$ at time slot $t$, which is bounded by a finite number, i.e. $\theta_e(t) \leq \mu_{max}$. 
 For any link $e\in E$, we assume that $\{\theta_e(t)\}_{t\geq0}$ are independent and that the mean service rate $\mu_e(t) = \EE[\theta_e(t)]$ may vary over time. 
 Additionally, we requires  the mean service rate to be lower bounded by a strictly positive constant, i.e. $\mu_e(t) \geq \mu_{min} > 0$; this assumption is also often imposed by the literature on optimal control of  queueing systems with time-varying statistics \cite{bipartite_queueing1, bipartite_queueing2}.
 The effective service rate of link $e$ at time slot $t$ is then given by:
\begin{align}
    b_e(t) = x_e(t) \theta_e(t), \label{eff_service1}
\end{align}
which characterizes the achievable data rate of the link.

Let $Q_e(t)$ be the physical queue of backlogged packets at link $e\in E$ that are waiting to be transmitted at the end of time slot $t$. Since any link $e$ receives $a_e(t)$  packet arrivals and can serve at most $b_e(t)$ packets during a time slot, the queueing dynamics evolves as:  
\begin{align}
\label{queue_dynamics1}
    Q_e(t+1) = \big(Q_e(t) + a_e(t) - b_e(t) \big)^+, \quad \forall e\in E,
\end{align}
where $[x]^+ = \max\{x, 0\}$.

In order to capture the realistic characteristics of modern wireless network, we incorporate the following requirements in our model:
\begin{itemize}
    \item \textit{Partial Observability:} For every link $e\in E$, both the instantaneous packet arrivals $a_e(t)$ and link's service capacity $\theta_e(t)$ are not available at the start of the time slot $t$ and thus cannot be used for making the scheduling decisions. At the end of time slot $t$, however, the nodes can accumulate statistics of the past time slot to obtain the packet arrivals $a_e(t)$'s and the service capacities of the activated links, i.e. those $\theta_e(t)$'s such that $x_e(t)=1$. For  unactivated links $e$ where $x_e(t) = 0$, though the information of $\theta_e(t)$ is not revealed, the effective service rate is $b_e(t) = 0$. Thus, given the knowledge of $a_e(t)$ and $b_e(t)$, the queuing dynamics \eqref{queue_dynamics1} for the next time slot $t+1$ can always be evaluated at the end of time slot $t$. The sequence of events within time slot $t$ is depicted in Figure \ref{fig: event_sequence}.
    \item \textit{Non-Stationary Dynamics:} We assume that both the mean packet arrivals $\lmb(t)=(\lambda_e(t))_{e\in E}$ and service rates $\muu(t)= (\mu_e(t) )_{e\in E}$  vary over time, i.e. non-stationary.
    \item \textit{Unknown Statistics:} All the statistics $\{ \lmb(t) \}_{t\geq 0}$ and $\{ \muu(t) \}_{t\geq 0}$ are unknown to the scheduler for making control decisions.
\end{itemize}
\subsection{Asymptotic Relationships and Notations}

Let $\QQ(t)=( Q_e(t) )_{e\in E}$, $\aA(t) = ( a_e(t) )_{e\in E}$, $\thetav(t) =( \theta(t) )_{e\in E} $ and $\bB(t) = (b_e(t) )_{e\in E}$ be respectively the vector of  queue lengths, packet arrivals, service capacities and effective service rates. For any two real numbers $x$ and $y$, we let $x \vee y = \max\{x, y\}$ and $x \wedge y = \min\{x, y\}$. For any vector $\xX = ( x_i)$ of real numbers and $p \in [1 , \infty)$, we denote $\|\xX \|_p$ as its $\ell_p$-norm. For the two cases of $p \in \{1, \infty\}$ used in this paper, we  have $\|\xX\|_1 = \sum_i |x_i|$ and $\|\xX\|_\infty = \max_i \{|x_i|\}$.
For two  positive multivariate functions $f(\xX)$ and $g(\xX)$, their asymptotic relationships \cite{intro_algo} are given in Table \ref{tab:asymptotic}.

\begin{table}
  \caption{Asymptotic Relationships}
  \label{tab:asymptotic}
  \begin{tabular}{ll}
    \midrule
    $f(\xX) = O(g(\xX))$ & there exists constants $M$ and $C$ such that  \\
     & $|f(\xX)| \leq  C |g(\xX)|$ for all $\xX$ with $\|\xX\|_\infty > M$.\\
     \hline 
    $f(\xX) = \Omega(g(\xX))$ & there exists constants $M$ and $C$ such that  \\
     & $|f(\xX)| \geq  C |g(\xX)|$ for all $\xX$ with $\|\xX\|_\infty > M$.\\
     \hline 
     $f(\xX) = \Theta(g(\xX))$ & $f(\xX) = O(g(\xX))$ and $f(\xX) = \Omega(g(\xX))$.\\
     \hline 
     $f(\xX) = o(g(\xX))$ & for every $\varepsilon >0$, there exists constant $M$ \\
     & such that $|f(\xX)| \leq  \varepsilon |g(\xX)|$ for all $\xX$ with   \\
     & $\|\xX\|_\infty > M$. In this case, we alternatively \\
     &  say $f(\xX)$ is sub-linear in $g(\xX)$.\\
    \bottomrule
  \end{tabular}
\end{table}


\subsection{Policy Space and Problem Statement}

For any variable affected by the control of the scheduling decisions, we add the superscript $\pi$ to acknowledge that it is under the action of the policy $\pi$. An \emph{admissible} policy $\pi$ at every time slot $t$ generates a scheduling decision $\xX^\pi(t) \in \Mm$ using only the knowledge of the past packet arrivals $\aA(0), \aA(1), ..., \aA(t-1)$, the past effective service rates $\bB(0), \bB(1), ..., \bB(t-1)$, and the past decisions $\xX(0), \xX(1),..., \xX(t-1)$ up to time $t-1$. Additionally, we consider \emph{idealized} policies, the definition of which is similar to that of an admissible policy except that at time slot $t$, it also has the full knowledge of the network statistics 
$\lmb(t)$ and $\muu(t)$
and can use them in making the scheduling decisions. 
The set of all admissible policies and  the set of all idealized polices are respectively denoted by $\Pi$ and $\Pi_s$. Under the simplified model whereby the network dynamics are  stationary, 
\cite{STAHLBUHK2019131}  designed a joint learning and scheduling algorithm in $\Pi$ that supports the same stability  region, i.e. the set of arrival rates under which the system is stabilizable, as that of idealized policies in $\Pi_s$. 
Nevertheless, generalization to the case of non-stationary network dynamics is non-trivial due to the analytical intractability of the capacity region. Moreover, the only previous work \cite{bipartite_queueing2} that attempts to learn non-stationary network dynamics  could achieve only a reduced stability region that is constrained by the window-based dynamics, thereby being sub-optimal. 

In this paper, we aim to develop a control scheme for the class of policies in $\Pi$ that maximizes the stability region of the network under our considered setting.



\section{Notion of Stability for Non-Stationary Network with Unknown Statistics}
\label{sec:notion_stability}

One main challenge of non-stationary network control is that the analytical characterization of the capacity region for the case of stationary network may no longer hold under the non-stationarity. In this Section, we propose a new notion of throughput-optimality that is more suitable to  the non-stationary  setting. For the simplified case of stationary network, we further establish the equivalence between our new notion and the conventional notion of throughput-optimality.  

\subsection{Assumption on Non-Staionary Dynamics}

For any  $t_1 < t_2$, we denote the total variation of the mean service rate by:
\begin{align}
    \gamma(t_1, t_2) = \sum_{t= t_1+1}^{t_2} \|\muu(t) - \muu(t-1) \|_\infty, \label{formula_variation1}
\end{align}
and stipulate the following mild assumption on the non-stationarity of the mean service rates.
\begin{assumption}
\label{vari_assumption}
For any $t_1 < t_2$, the total variation is upper-bounded by  $ \gamma(t_1, t_2) = O(|t_2-t_1|^\alpha)$ for some  $\alpha\in [0, 1)$.
\end{assumption}
Our  assumption only  requires the total variation of mean service rates over any time period to grow sub-linearly in  time, thereby ensuring that the network dynamics do not vary too aggressively. Similar  assumptions have been extensively used the literature of learning in non-stationary environments \cite{combMAB_nonstationary1, combMAB_nonstationary2, combMAB_nonstationary3}.


\subsection{Performance Metrics}
\label{sec:performance_metrics}

Before characterizing the stability regions of interest, we first define the  $Q_T$ measure that captures the growth of queue size in expectation, and present the equivalent definition of mean rate stability \cite{neelybook1} under  the $Q_T$ measure.

\begin{definition}[$Q_T$ Measure] The total expected queue length at time $T$ under a control of policy $\pi$ is quantified by $ Q^\pi_T = \EE\big[\sum_{e\in E} Q^\pi_e(T)\big]$.
    
\end{definition}


\begin{definition}[Mean Rate Stability] 
A network is mean rate stable  under a  policy $\pi$ if:
\begin{align*}
    \lim_{T\to \infty} \frac{\EE[\sum_{e\in E} Q^\pi_e(T)]}{T} = 0, 
\end{align*}
or equivalently $Q_T^\pi = o(T)$.
\end{definition}


The notion of stability region of a  policy $\pi$   describes the set of arrival rate vectors such that mean rate stability could be achieved under $\pi$.  The stability region $\Lmb$ is the  region that can be achieved by the  class $\Pi$ of admissible policies, as formally defined below.

\begin{definition}[Stability Region] The stability region of the class of admissible policies is defined as:
    \begin{align*}
    \Lmb = \big\{ \{\lmb(t)\}_{t\geq 0}:  \exists \pi \in \Pi \text{ such that } Q_T^\pi = o(T) \big\}.
\end{align*}
\end{definition}


Similarly, we define the idealized stability region $\Lmb_s$ of the class $\Pi_s$ of idealized policies.

\begin{definition}[Idealized Stability Region] The stability region of the class of idealized policies is defined as:
    \begin{align*}
    \Lmb_s = \big\{ \{\lmb(t) \}_{t\geq 0}:  \exists \pi \in \Pi_s \text{ such that } Q_T^\pi = o(T) \big\}.
\end{align*}
\end{definition}

Note that we always have $\Lmb \subseteq \Lmb_s$, since the idealized policies have the full network statistics as opposed to the admissible policies. For the simplified model of stationary dynamics, a special case of our setting where Assumption \ref{vari_assumption} trivially holds with  $\alpha = 0$, \cite{STAHLBUHK2019131}  designed an admissible policy  stabilizing any arrival rate vector in $int(\Lmb_s)$\footnote{\cite{STAHLBUHK2019131}  considers maximal matching for scheduling under matching constraint, and thus achieves the interior of $\frac{1}{2} \Lmb_s$ as the stability region.  However, by replacing maximal matching with maximum matching, we can achieve the full stability region $int(\Lmb_s)$.}, i.e. the interior of the idealized stability region. The algorithmic development and analysis of  \cite{STAHLBUHK2019131} heavily rely on the fact that under the stationary dynamics whereby $\lmb = \lmb(t)$ and $\muu = \muu(t)$ for all $t\geq 0$, the idealized stability region $\Lmb_s$ can be further characterized  by  the existence of a   policy $\pi^* \in \Pi_s$  such that:
\begin{align}
    \lambda_e = \lim_{T\to \infty} \frac{1}{T} \sum_{t=0}^{T-1} a_e(t) \leq  \lim_{T\to \infty} \frac{1}{T} \sum_{t=0}^{T-1} b^{\pi^*}_e(t), \text{ } \forall e \in E. \label{stationary_equation}
\end{align}
However, under  non-stationarity, the above limits may not even exist, thereby hindering the adoption of stability region's characterization as in the case of stationary dynamics. Such  analytical intractability of the stability region  is central to the problem of optimal control for  non-stationary network \cite{network_nonstationary1, adversarial_control1}.

\subsection{Notion of Throughput-Optimality for Non-Stationary Network}


In this Section, we propose a novel notion of throughput-optimality for non-stationary networks. 
For the simplified setting of stationary network, the conventional notion of stability  defines a policy $\pi$ to be throughput-optimal if it can stabilize the system for any  arrival rate $\lmb \in int(\Lmb_s)$, i.e. in the interior of the idealized stability region. This is equivalent to $\lmb \in (1-\varepsilon)\Lmb_s$ for some $\varepsilon > 0$, which is then usually incorporated with the analytical characterization of $\Lmb_s$ for establishing the stability of MaxWeight-type algorithms. However, in the context of non-stationary networks, such an approach may not be directly applicable due to the analytical intractability of the idealized stability region $\Lmb_s$. 
To this end, we first present our new definition of approximate stability region, which is central to  our  throughput-optimality notion and algorithmic development.

\begin{definition}[Approximate  Stability Region] Given any $\beta \in [0, 1]$, we define the approximate region $\Lmb_s(\beta)$ as:
    \begin{align*}
    \Lmb_s(\beta) = \big\{ \{\lmb(t) \}_{t\geq 0}:  \exists \pi \in \Pi_s \text{ such that } Q_T^\pi = O\big(T^\beta \big) \big\}
\end{align*}
\end{definition}
We now derive the key properties of $\Lmb_s(\beta)$ and its relation to the idealized stability region $\Lmb_s$ in the following  Lemma, whose proof is deferred to Appendix \ref{appen_approx_region1}.

\begin{lemma}
\label{lma:approx_region1}
 The set $\Lmb_s(\beta)$ is expanding for increasing $\beta$, i.e. if $0\leq  \beta_1 \leq \beta_2 \leq  1$, then  $ \Lmb_s(\beta_1) \subseteq  \Lmb_s(\beta_2)$. 
 Moreover, for any $\beta \in [0, 1)$, we have  $ \Lmb_s(\beta) \subseteq \Lmb_s  \subseteq \Lmb_s(1) $.
\end{lemma}

Lemma \ref{lma:approx_region1} suggests that the  region $\Lmb_s(\beta)$ grows arbitrarily close to $\Lmb_s$ as $\beta$ approaches $1$. Moreover, leveraging this notion of approximate stability region, the next Theorem establishes the characterization of the true stability region $\Lmb$.

\begin{theorem}
\label{thm:cap_region1}
Under Assumption \ref{vari_assumption}, we have the following characterization of the stability region:
\begin{align*}
     \Lmb_s(\beta) \subseteq \Lmb \subseteq  \Lmb_s,
\end{align*}
 for any $\beta \in [0, 1)$.
\end{theorem}
\begin{proof}
 As  the idealized policies  have the full network statistics as opposed to the admissible policies, the idealized stability region $ \Lmb_s$ trivially subsumes the stability region $ \Lmb$. 
The proof of $ \Lmb_s(\beta) \subseteq \Lmb$ is based on our development of the admissible policy MW-UCB in Section \ref{sec:main_algo} that, given any $\beta \in [0, 1)$, achieves mean rate stability for any set of arrival rates $(\lmb(t))_{t\geq 0} \in \Lmb_s(\beta)$ (Theorem \ref{thrm:througput_optimality}).
\end{proof}

Motivated by Theorem \ref{thm:cap_region1}, we propose the following notion of throughput-optimality for non-stationary network control.

\begin{definition}[Throughput-Optimality]
\label{def:throughput}
 A policy $\pi\in \Pi$ is throughput optimal if given any $\beta \in [0, 1)$, the network under $\pi$ is mean rate stable for any $\{ \lmb(t) \}_{t\geq 0} \in \Lmb_s(\beta)$.
\end{definition}

Under the above definition,  we aim to develop an admissible policy in $\Pi$ that is throughput-optimal for our considered setting of non-stationary and partially-observable network. 


\subsection{Connection to Traditional  Notion of Throughput-Optimality for Stationary Networks}

We further demonstrate the equivalence of our throughput-optimality notion to the usual notion in the case of stationary network whereby $\lmb = \lmb(t)$ and $\muu = \muu(t)$ for all $t\geq 0$. As discussed in Section \ref{sec:performance_metrics}, the idealized stability region $\Lmb_s$ can be characterized by:
\begin{align*}
    \bar{\Lmb}_s = \big\{ \lmb:  \exists \pi^* \in \Pi_s \text{ such that $\pi^*$ satisfies \eqref{stationary_equation}}\big\},
\end{align*}
and an admissible policy is throughput-optimal (in the usual notion) if it can stabilize any arrival rate $\lmb \in  \bar{\Lmb}_s$\footnote{Here, we use $ \bar{\Lmb}_s$ to emphasize that this is a special case of ${\Lmb}_s$ where  such characterization based on \eqref{stationary_equation} only holds for the stationary network setting.}.  The next Theorem illustrates that our new notion of throughput-optimality implies the usual notion of throughput-optimality in the stationary network control problem. 

\begin{theorem}
\label{thrm:equivalence}
 Under the stationary network setting, if a policy $\pi \in \Pi$ is throughput-optimal according to Definition \ref{def:throughput}, then the network under $\pi$ is mean rate stable for any $\lmb \in \bar{\Lmb}_s$.
\end{theorem}

\textsc{Proof Sketch.} 
Given $\lmb \in \bar{\Lmb}_s$, we can show via Lyapunov drift analysis that the Max-Weight (MW) policy with full knowledge of the statistics achieves $Q_T^{\textsc{MW}} = O\big(T^{\frac{1}{2}} \big)$. Consequently, this implies that $\lmb \in \Lmb_s(\frac{1}{2})$. Since by our Definition  \ref{def:throughput}, a throughput-optimal policy $\pi$ would support the stability region $\Lmb_s(\frac{1}{2})$, it thus guarantees mean rate stability for any $\lmb \in \bar{\Lmb}_s$. The  full proof is given in Appendix \ref{appen_equivalence}. 


\section{Scheduling with non-stationary and unknown channel statistics}
\label{sec:main_algo}

In this Section, we present  MW-UCB as a provably throughput-optimal policy for non-stationary network control. 
We  provide the preliminaries of Upper-Confidence Bound (UCB) for learning uncertain channel statistics in Section \ref{sec:ucb_preliminaries} and  the algorithmic development of MW-UCB in Section \ref{sec:alg_develop}. The throughput-optimality and stability results of MW-UCB then follow in Section \ref{sec:theoretical_results}.

\subsection{Upper-Confidence Bound (UCB)  for Learning Uncertain Channel Statistics}
\label{sec:ucb_preliminaries}
Central to our problem is the learning of not only the unknown links' service rates but also the scheduling decisions that can maximize the overall network's throughput. We start by considering a simplistic problem setting in which the network dynamics are stationary and the objective is to attain the maximum possible total service capacity (in expectation) of the network, and show that a simple  Upper-Confidence Bound (UCB) algorithm is close-to-optimal in this scenario.  
However, while having the potential for being the solution for network control under uncertain channel statistics, the UCB algorithm in its original form  lacks the adaptivity to deal with non-stationary dynamics and is a pure learning scheme in nature, which is not designed to deal with sophisticated control tasks as in our original problem.

\subsubsection{A Simplistic Problem Setting and Application of UCB}
At any time slot $t$, the scheduling decision $\xX(t)$ yields  in expectation the service of $\mu_e(t) x_e(t)$  for link $e \in E$ and thus the total service of:
\begin{align}
    \sum_{e\in E} \mu_e(t) x_e(t). \label{simple_ucb1}
\end{align}
Now, we turn into a simplified objective of maximizing the total service \eqref{simple_ucb1} over the time horizon $T$ of the network and further assume stationary dynamics of the links' service rates, i.e. $\mu_e(t) = \mu_e, \forall e\in E, t\geq 0$. Under this setting, an idealized policy with full knowledge of the  statistics $\muu= ( \mu_e)_{e\in E}$ would make the scheduling decision at every time slot $t$ that maximizes \eqref{simple_ucb1}, i.e.
\begin{align}
    \xX(t) = \xX^* =  \argmaxE_{\xX \in \Mm} \big\{ \sum_{e\in E} \mu_e x_e \big\}. \label{simple_ucb2}
\end{align}
However, such statistics $\muu= ( \mu_e)_{e\in E}$ are unknown in practice and thus must be learned, under our requirement of partial observability, via samples of service capacities of links having been activated. This gives rise  to the exploration/exploitation tradeoff, where the controller must simultaneously learn the channel statistics $\muu$ and utilize the existing information of observed service capacities to schedule transmissions achieving high total throughput. In particular, the problem of solving \eqref{simple_ucb2} over the time horizon $T$ can be characterized as   combinatorial multi-armed bandit (CMAB) with linear reward in stationary environment, which can be addressed by the class of  UCB algorithms \cite{mab_origin1, yigai1, cmab1}. We  hereby consider the UCB algorithm  in \cite{cmab1}, which proceeds as follows. At any time slot $t$ and for every edge $e \in E$, the UCB algorithm keeps track of $T_t(e)$   and $\hat{\theta}_e(t)$, which respectively correspond to the number of times link $e$ has been activated and observed up to time $t$, and the empirical mean of all the observations of the service capacities, i.e. those $\theta_e(s)$ such that $x_e(s) = 1$ for $s\in [0, t)$. The UCB weights are computed according to:
\begin{align*}
    U_e(t) =  \hat{\theta}_e(t) + \sqrt{\frac{3 \log(t)}{2T_t(e)}}, \forall e\in E,
\end{align*}
which are then used for constructing the scheduling decision as:
\begin{align}
     \xX(t)  =  \argmaxE_{\xX \in \Mm} \big\{ \sum_{e\in E} U_e(t) x_e \big\}. \label{simple_ucb3}
\end{align}
The total  difference in achievable total expected service capacity between  the   maximizing policy with the full knowledge of the statistics $\muu$ that makes the decision $\xX^*$ as in \eqref{simple_ucb2} and the UCB algorithm that makes the decision $\xX(t)$ as in \eqref{simple_ucb3} is captured by the  regret:
\begin{align}
    \Rr_0(T) = T \sum_{e\in E} \mu_e x^*_e - \sum_{t= 0}^{T-1} \sum_{e\in E} \mu_e x_e(t). 
\end{align}
This type of metrics is also used by \cite{STAHLBUHK2019131} for characterizing the performance and exploration/exploitation tradeoff of their joint learning and scheduling algorithm.
From  \cite[Theorem 5]{cmab1}, we have $\Rr_0(T) = O\big(\log(T)\big)$, which guarantees only logarithmic growth  in  total error if the UCB algorithm is applied. Moreover, this regret bound is asymptotically tight \cite[Proposition 1]{cmab1}.  






\subsubsection{Limitations of The Conventional UCB Algorithm}
While attaining competitive performance in the simplistic problem setting, the conventional UCB algorithm lacks the generality to  readily be extended to deal with our problem  of interest. First, vanilla UCB is known to be inappropriate for handling non-stationary dynamics  \cite{ucb_fail1}. Second, the formulation \eqref{simple_ucb2} that permits the adoption of UCB as a direct solution   does not take into account the control of the system  under arbitrary arrival rates: for example, if a policy aims to attain the maximum possible total service and hence  always makes the scheduling decision $\xX^*$ as in \eqref{simple_ucb2}, it would inevitably overload certain unactivated links, i.e. $e\in E$ with $x^*_e = 0$, to which there are packet arrivals over time. On the other hand, the  Max-Weight policy \cite{neelybook1, BP1} that incorporates the queue lengths into making scheduling decision can adapt to the dynamics of   arbitrary  arrival rates.
Consequently, the solution for scheduling in non-stationary wireless networks with partial observability and unknown statistics   requires  the interplay between learning and network control. In the next Section, we present our main algorithm that combines the Max-Weight policy with UCB to address these aforementioned  challenges through its joint learning and scheduling scheme.


\subsection{The MW-UCB Algorithm}
\label{sec:alg_develop}


We proceed to develop our  scheduling algorithm, termed MW-UCB, based on a frame-based variant \cite{STAHLBUHK2019131} of the Max-Weight policy \cite{neelybook1, BP1} and the augmentation of the sliding-window UCB \cite{combMAB_nonstationary1} in the weight construction for adaptively learning the channels’ statistics under non-stationarity. 
The full MW-UCB policy is depicted in Algorithm \ref{alg:MW_UCB} with the convention that $0/0=0$. 

\begin{algorithm}
\caption{Max-Weight with UCB (\textbf{MW-UCB})}
\label{alg:MW_UCB}
\KwIn{ graph $\mathcal{G}(V, E)$, restart period $\tau$, window size $d \leq \tau$ }
\For{$t=1, ..., T$}{
\If{$t = \tau_j \in \mathcal{T}=\{\tau_0, \tau_1,..., \tau_{K}\}$}{
  Initialize $\phi_e(\tau_j)=0, N_e(\tau_j) = 0, \hat{\mu}_e(t) = 0, \forall e \in E$ \\
  Reset the weights: $w_e(\tau_j) = \frac{Q_e(\tau_j)}{ \| \QQ(\tau_j) \|_\infty },  \forall e \in E$ \label{reset_weight1}\\
}
\If{$t \in (\tau_j, \tau_{j+1})$}{
\For{$e\in E$}{
$\phi_e(t) = \phi_e(t-1)  -   \mathbbm{1}_{\{t \geq \tau_j + d \}}  b_e(t-d)  + b_e(t-1)  $ \label{serving_cap2} \\
$N_e(t) =  N_e(t) -    \mathbbm{1}_{\{t \geq \tau_j + d\}} x_e(t-d) +  x_e(t-1)  $ \label{times_activated2} \\
$\hat{\mu}_e(t) = \frac{\phi_e(t)}{N_e(t)} $ \textcolor{blue}{  $\triangleright$ empirical estimate of $\mu_e(t)$} \label{empi_ave1}\\
$w_e(t) = w_e(\tau_j) $ \textcolor{blue}{  $\triangleright$ fixing queue lengths} \label{reset_weight2}\\
}
}
$\rho_e(t) = \sqrt{ \frac{3 \log(\tau)}{ 2 N_e(t) } } \text{ (or $\infty $ if $N_e(t) = 0$)}, \forall e\in E$ \\
$\bar{W}_e(t) = \min\{ w_e(t)  \hat{\mu}_e(t) + \rho_e(t), 1\} , \forall e\in E $ \textcolor{blue}{$\triangleright$  computing UCB weights} \label{ucb_weight1}\\

\textbf{[Scheduling]} Activate the link activation vector: 
$\xX(t) = \argmaxE_{\xX \in \Mm} \big\{ \sum_{e\in E} \bar{W}_e(t) x_e \big\}$ \label{mw_scheduling1}\\
Observe $\aA(t)$ and $\bB(t)$, and update the queues: $Q_e(t+1) = \big(Q_e(t) + a_e(t) - b_e(t) \big)^+, \quad \forall e\in E,$
}
\end{algorithm}

For the class of idealized policies in $\Pi_s$, the Max-Weight policy \cite{neelybook1} that at  time slot $t$ weights each edge $e$ by $Q_e(t) \mu_e(t)$  and consequently schedule the link activation vector according to:
\begin{align}
    \xX(t) = \argmaxE_{\xX \in \Mm} \big\{ \sum_{e\in E} Q_e(t) \mu_e(t) x_e \big\} \label{MW_scheduling}
\end{align}
is known to be throughput-optimal for the case of stationary network  \cite{STAHLBUHK2019131, neelybook1} and obtains competitive performance on adversarial network control  \cite{adversarial_control1}. Nevertheless, under our considered model,  the vector $\muu(t)$ of mean service rates is unknown a priori, thereby hindering any direct adoption  of the Max-Weight policy. The algorithmic design for joint network control and learning of the weights $\{Q_e(t) \mu_e(t) \}_{e\in E}$ faces two challenges. First, $Q_e(t) \mu_e(t)$ is time-varying due to the dynamics of the queue length $Q_e(t)$ and the non-stationarity of $\mu_e(t)$. Second, the evolution of the weight $Q_e(t) \mu_e(t)$ is coupled with the scheduling decision $\xX(t)$ due to its interdependence with effective service rate via \eqref{eff_service1} and thus the queueing dynamics \eqref{queue_dynamics1}. To address these challenges, we  periodically freeze the queue length information in the weight instantiation, which  helps to alleviate a source of non-stationarity and decouples the weight evolution from the scheduling decision. 
Specifically, our method partitions the time horizon $T$ into frames of size $\tau$, where the $j^{th}$ frame  begins at time slot $\tau_j = j \tau$, called restart point. We allow the last frame to have size potentially less than $\tau$ and let $\mathcal{T}=\{\tau_0, \tau_1,..., \tau_{K} \}$ be the set of all  restart points, i.e. $K$ is the largest number such that $\tau_{K} < T$. Then for any $t\in [\tau_j, \tau_{j+1})$, we use  the  normalized queue backlogs at the restart point $\tau_j$ as the unified weights (Line \ref{reset_weight1} and Line \ref{reset_weight2} of  Algorithm \ref{alg:MW_UCB}):
\begin{align}
    w_e(\tau_j) = \frac{Q_e(\tau_j)}{ \| \QQ(\tau_j) \|_\infty },  \forall e \in E, \label{normalized_weight1}
\end{align}
and aim to solve the following "relaxed" problem of   \eqref{MW_scheduling} with simplified time-varying weight structure:
\begin{align}
    \xX(t) &= \argmaxE_{\xX \in \Mm} \big\{ \sum_{e\in E} Q_e(\tau_j) \mu_e(t) x_e \big\} \label{MW_scheduling_relaxed}\\
    &\overset{\eqref{normalized_weight1}}{=} \argmaxE_{\xX \in \Mm} \big\{ \sum_{e\in E} w_e(\tau_j) \mu_e(t) x_e \big\}. \label{MW_scheduling_relaxed2} 
\end{align}
Note that the objective in \eqref{MW_scheduling_relaxed}  is the "approximation" of the objective in \eqref{MW_scheduling} with error growing linearly in $\tau$. 
Moreover, the problem of solving \eqref{MW_scheduling_relaxed2} over $\tau$ time slots from $\tau_j$ to $\tau_{j+1}-1$ can be characterized as stochastic combinatorial multi-armed bandit (SCMAB) problem in non-stationary environment, whereby the mean reward of each arm, i.e. $w_e(\tau_j) \mu_e(t)$, varies over time and is independent of the action, i.e. scheduling decision. To this end, we adopt the combinatorial UCB with sliding window (CUCB-SW) algorithm \cite{combMAB_nonstationary1} for dealing with SCMAB under non-stationarity.
Specifically,  CUCB-SW  is restarted at the beginning of each frame with the newly updated queue weights for the joint learning of the mean service rates $\muu(t)$'s and control of the system. Given the sliding window of size $d$ as a hyper-parameter to be chosen later, the algorithm computes the  estimate $\hat{\mu}_e(t)$ of the true mean service rate $\mu_e(t)$ as the local empirical average of the observed service capacities in the last $d$ time slots. Formally, for $t\in [\tau_j, \tau_{j+1})$, i.e. within the $j^{th}$ frame, and any $e\in E$, the following quantities:
\begin{align}
    &\phi_e(t) = \sum_{s = \tau_j \vee (t - d) }^{t-1} \mathbbm{1}_{\{x_e(s) = 1 \}}  \theta_e(s) = \sum_{s =  \tau_j \vee (t - d) }^{t-1}  b_e(s), \label{serving_cap1} \\
       &N_e(t)  =  \sum_{s = \tau_j \vee (t - d) }^{t-1} \mathbbm{1}_{\{x_e(s) = 1 \}} =  \sum_{s = \tau_j \vee (t - d) }^{t-1} x_e(s), \label{times_activated1}
\end{align}
respectively denote the total \emph{observed} service capacities of link $e$ and the number of times it had been \emph{activated} up to time $t-1$ over the last $d$ time slots. 
Line \ref{serving_cap2} and Line \ref{times_activated2} of  Algorithm \ref{alg:MW_UCB} equivalently rewrite \eqref{serving_cap1} and \eqref{times_activated1}  in recursive forms for actual iterative updates in the algorithm. 
Then the local empirical average can be computed accordingly via $\hat{\mu}_e(t) = {\phi_e(t)}/{N_e(t)}$ as in Line \ref{empi_ave1}. Finally, the UCB weights are computed by (Line  \ref{ucb_weight1}): 
\begin{align*}
    \bar{W}_e(t) = \min\bigg\{ w_e(\tau_j)  \hat{\mu}_e(t) +  \sqrt{ \frac{3 \log(\tau)}{ 2 N_e(t) }}   , 1 \bigg\} , \forall e\in E, 
\end{align*}
to be  used for  constructing of the scheduling decision  (Line \ref{mw_scheduling1}) as:
\begin{align}
    \xX(t) = \argmaxE_{\xX \in \Mm} \big\{ \sum_{e\in E} \bar{W}_e(t) x_e \big\}. \label{mw_ucb_scheduling23}
\end{align}

In order to capture the loss due to learning when CUCB-SW is applied to solve \eqref{MW_scheduling_relaxed2} within the $j^{th}$ frame, we consider the regret: 
\begin{align}
\nonumber
   \Rr(\tau_j) &=  \sum_{t= \tau_j}^{\tau_{j+1}-1} \max_{\xX\in \Mm} \sum_{e\in E} w_e(\tau_j)  \mu_e(t)  x_e \\
   &\quad - \sum_{t= \tau_j}^{\tau_{j+1}-1}\sum_{e\in E}  w_e(\tau_j) \mu_e(t)   x_e(t), \label{original_regret0}
\end{align}
which characterizes the gap in objective between  the maximizing policy with the full knowledge of the statistics $\muu(t)$ for $t \in [\tau_j, \tau_{j+1})$ that solves \eqref{MW_scheduling_relaxed2} and the considered policy that solves \eqref{mw_ucb_scheduling23}.
The following Lemma \ref{lma:regret_bound}, whose proof is given in  Appendix \ref{append_drift_bound1}, provides an upper bound for $\Rr(\tau_j)$.

\begin{lemma}
\label{lma:regret_bound} 
Under MW-UCB,  the regret $\Rr(\tau_j)$ can be bounded by:
\begin{align*}
    \Rr(\tau_j) &\leq |E|\bigg( \frac{\tau}{d} +1\bigg) \bigg(2\sqrt{6 \log(\tau)} + 48 \sqrt{d} \log(\tau)  \bigg) \\
    &\quad +4 |E|  d \cdot \gamma(\tau_j, \tau_{j+1})  + |E|   \frac{\tau}{\sqrt{d}}+ \frac{\pi^2}{3} |E|^2 \mu_{max} \\
    &\quad+ \frac{\pi^2}{6} |E|^2 \mu_{max}    \log\big(  2 d^{1/2}  \big). 
\end{align*}
Under  Assumption \ref{vari_assumption}, which gives $\gamma(\tau_j, \tau_{j+1})  = O(\tau^\alpha)$, and by setting  $d= \Theta\big(\tau^{\frac{2}{3}(1-\alpha)}\big)$, we further have  $\Rr(\tau_j) = O\big( \log(\tau) \tau^{\frac{1}{3} (\alpha + 2)}  \big)$.
\end{lemma}
The guarantee in Lemma \ref{lma:regret_bound} demonstrates that the average loss due to learning over $\tau$ time slots of the $j^{th}$ frame vanishes as $\tau$ grows in the sense that $\frac{\Rr(\tau_j)}{\tau} \to 0$ as $\tau \to \infty$ under our mild assumption on the learnability of the system. This is crucial for establishing the throughput-optimality of MW-UCB in the next Section.


\subsection{Throughput-Optimality and Stability Results}
\label{sec:theoretical_results}

In this Section, we prove the throughput-optimality of MW-UCB and, as a byproduct, its strong stability in a  region constrained by window-based dynamics. 
The key components of the proof leverage the regret bound for learning non-stationarity (Lemma \ref{lma:regret_bound}) in the analysis of the frame-based Lyapunov drift and  non-trivially generalize the shedding technique  from the adversarial network control literature \cite{adversarial_control1}. In particular, to address the analytical intractability of the capacity region, we shed the traffic of the original system to obtain a new imaginary system whose traffic is within the window-based region, which is formally described in Definition \ref{def_window}. While the shedding process incurs  additional term in the  queue bound as a tradeoff, it makes the imaginary network dynamics tractable, from which stability of MW-UCB can be derived.  





\begin{definition}[Window-Based Region]
\label{def_window}
The window-based region $\Cc(W, \varepsilon)$ of the class of idealized policies is defined as:
\begin{align*}
    \Cc(W, \varepsilon) =\big\{ &\{\lmb(t) \}_{t\geq 0}:  \exists \pi \in \Pi_s \text{ such that for $q= 0, W, 2W,...$} \\
     & \sum_{t=q}^{q+W-1} \lambda_e(t) \leq (1-\varepsilon) \sum_{t=q}^{q+W-1} \EE[b^\pi_e(t)], \text{ } \forall e \in E \big\}.
\end{align*}
\end{definition}

Specifically, the sequence of arrival rates $\{\lmb(t) \}_{t\geq 0}$ satisfies the window-based region, parameterized by     window size $W$ and a shrinkage term $\varepsilon$, if there exists an idealized policy such that  the total mean  arrivals  are less than  a fraction $1-\varepsilon$ of the total mean services over a window of $W$ slots starting at every starting point $t= 0, W, 2W,...$ Next, we proceed to establish  the main Theorem on the  throughput-optimality of  MW-UCB. 


 


\begin{theorem}
\label{thrm:througput_optimality}
Under Assumption \ref{vari_assumption},  MW-UCB  is throughput-optimal. 
\end{theorem}
\begin{proof}

Given any $\beta \in [0, 1)$, we show that MW-UCB can achieve mean rate stability for any $\{\lmb(t) \}_{t\geq 0} \in  \Lmb_s(\beta)$.
We now consider an \emph{imaginary} system that is obtained by imitating the same link service process $\{\thetav(t) \}_{t\geq 0}$ as the original system's and shedding a certain amount of traffic from the original system's arrivals $\{\aA(t) \}_{t\geq 0}$ to obtain a new sequence of arrivals $\{\tilde{\aA}(t) \}_{t\geq 0}$ with $\tilde{\lambda}_e(t) = \EE[ \tilde{a}_e(t)], \forall e \in E, t\geq 0$. Denote the amount of shed traffic within the time horizon by:
\begin{align}
    X_T = \sum_{t=0}^{T-1} \sum_{e\in E} a_e(t)-\sum_{t=0}^{T-1} \sum_{e\in E} \tilde{a}_e(t). \label{XT}
\end{align}
For some $\varepsilon \in (0, 1)$ to be determined later, we consider the shedding scheme as in the following Lemma \ref{lma:shedding1}, whose proof is deferred to Appendix \ref{appen_shedding}. 
\begin{lemma}
\label{lma:shedding1}
Given $\{\lmb(t) \}_{t\geq 0} \in  \Lmb_s(\beta)$ and any $\varepsilon \in (0,1)$, there exists  a shedding procedure such that $\{\tilde{\lmb}(t) \}_{t\geq 0} \in  \Cc(\tau, \varepsilon)$ and:
\begin{align}
    \EE[X_T] = O\big( T^{\beta+1} \tau^{-1} + \varepsilon T +\tau \big). \label{shedding_cost}
\end{align}
\end{lemma}
Intuitively, Lemma \ref{lma:shedding1} suggests that, for analysis, despite the potential analytical intractability of the approximate region $ \Lmb_s(\beta)$, we can shed traffic to obtain an imaginary system that is constrained in $ \Cc(\tau, \varepsilon)$ and thus more tractable with the tradeoff as characterized by \eqref{shedding_cost}. 
When MW-UCB is applied to the original system, it produces the sequence of decisions $\{\xX^{\textsc{MW-UCB}}(t)\}_{t\geq 0}$.  
Let $\widetilde{\QQ}(t) = \{ \tq_e(t)\}_{e\in E}$ be the virtual queue length vector at time slot $t$ if such sequence of decisions $\{\xX^{\textsc{MW-UCB}}(t)\}_{t\geq 0}$ is applied to the imaginary system. Then the following Lemma \ref{lma:relate_to_imag} upper-bounds the $Q_T$ measure of MW-UCB in the original system.
The proof is given in Appendix \ref{appen_relate_to_imag}.
\begin{lemma}
\label{lma:relate_to_imag}
We have the following bound:
\begin{align}
   Q_T^{\textsc{MW-UCB}} \leq \EE[X_T] + \EE\big[ \sum_{e\in E} \widetilde{Q}_e(T)\big] . \label{relate_to_imag} 
\end{align}
\end{lemma}

In particular, though the shedding process incurs the term  $\EE[X_T]$, which can be bounded as in Lemma \ref{lma:shedding1}, in the queue bound of the original system,  we are now left with bounding the virtual queues $\WQQ(t)$ which evolve over the imaginary system that is more tractable.   
Next, we further provide guarantee for the total expected queue length of the imaginary system, i.e. the second term of \eqref{relate_to_imag}, as follows.
\begin{lemma}
\label{lma:imag_queue_bound}
Under Assumption \ref{vari_assumption} and given $\{\tilde{\lmb}(t) \}_{t\geq 0} \in  \Cc(\tau, \varepsilon)$, there exists some universal constant $c_1 > 0$ such that if $  c_1 \log(\tau) \tau^{\frac{1}{3}(\alpha -1) } \leq  \varepsilon < 1 $, MW-UCB with $d= \Theta\big(\tau^{\frac{2}{3}(1-\alpha)}\big)$ being applied to the imaginary system satisfies:
\begin{equation}
     \EE\big[ \sum_{e\in E} \widetilde{Q}_e(T)\big] = O\big( T^{\frac{1}{2}}\tau^{\frac{1}{2}} + T^{1+\frac{\beta}{2}}  \tau^{-\frac{1}{2}}  + \varepsilon^{\frac{1}{2}} T \big) .\label{imag_queue_bound}
\end{equation}
\end{lemma}
The proof of Lemma \ref{lma:imag_queue_bound} can be found in  Appendix \ref{appen_imag_queue_bound}. 
Notice that under our assumption $\alpha < 1$ (implying $\lim_{\tau \to \infty} \log(\tau) \tau^{\frac{1}{3}(\alpha -1) } = 0$), without loss of generality (WLOG), we  can consider $\tau$ large enough so that $c_1 \log(\tau) \tau^{\frac{1}{3}(\alpha -1)} < 1$. 
The requirement that the considered "shrinkage" $\varepsilon$ must be bounded away from $0$ by such quantity reflects the loss due to learning.  
Note that the whole shedding process only serves for analytical purposes, i.e. we can shed the original system into the new imaginary system in the sense of Lemma \ref{lma:shedding1} for any arbitrary $\varepsilon \in (0, 1)$. 
Finally, by setting $\varepsilon = c_1 \log(\tau) \tau^{\frac{1}{3}(\alpha -1)}$, we plug \eqref{shedding_cost} and \eqref{imag_queue_bound} into \eqref{relate_to_imag} to obtain the following bound for the $Q_T$ measure of MW-UCB with $d= \Theta\big(\tau^{\frac{2}{3}(1-\alpha)}\big)$:
\begin{align*}
    Q_T^{\textsc{MW-UCB}} &= O\big( T^{\beta+1}\tau^{-1} + T \sqrt{\log(\tau)} \tau^{\frac{1}{6} (\alpha -1) }  +T^{1+\frac{\beta}{2}}  \tau^{-\frac{1}{2}}  + T^{\frac{1}{2}}\tau^{\frac{1}{2}} \big).  
\end{align*}
Therefore, MW-UCB with  $\tau = T^{\frac{1}{2} (\beta + 1)}$ and $d= \Theta\big(\tau^{\frac{2}{3}(1-\alpha)}\big)$  can achieve:
\begin{align}
\nonumber
 Q_T^{\textsc{MW-UCB}} &= O\big( T^{\frac{1}{2} (\beta + 1) }  + \sqrt{\log(T)} T^{1- \frac{1}{12}(\beta+1)(1-\alpha) } +T^{ \frac{1}{4} (\beta+3) }   \big)    \\
 &= o(T), \label{througput_optimality_proof1}
\end{align}
where the last line holds because $\alpha, \beta \in [0, 1)$.
Since \eqref{througput_optimality_proof1} asserts the mean rate stability  of MW-UCB  for  $\{\lmb(t) \}_{t\geq 0} \in  \Lmb_s(\beta)$ given any $\beta \in [0, 1)$, we conclude that MW-UCB is throughput-optimal.
\end{proof}

Additionally, we derive the strong stability of MW-UCB for the window-based region in the following Corollary, whose proof is given in  Appendix \ref{appen_strong_stability}. 

\begin{corollary}
\label{cor:strong_stability}
Under Assumption 1, MW-UCB with a fixed window size $\tau$ and $d= \Theta\big(\tau^{\frac{2}{3}(1-\alpha)}\big)$ achieves strong stability for any  $\{{\lmb}(t) \}_{t\geq 0} \in  \Cc(\tau, \varepsilon)$ with any $\varepsilon >  c_1 \log(\tau) \tau^{\frac{1}{3}(\alpha -1) }$ for some universal constant $c_1 > 0$, i.e.
\begin{align}
    \limsup_{T\to \infty} \frac{1}{T} \sum_{t=0}^{T-1} \sum_{e\in E}  \EE[ Q_e(t)] < \infty. 
\end{align}
\end{corollary}


\section{Numerical Simulation}
\label{sec:exp}

\begin{figure}[H]  
\centering
\includegraphics[width=0.27\textwidth]{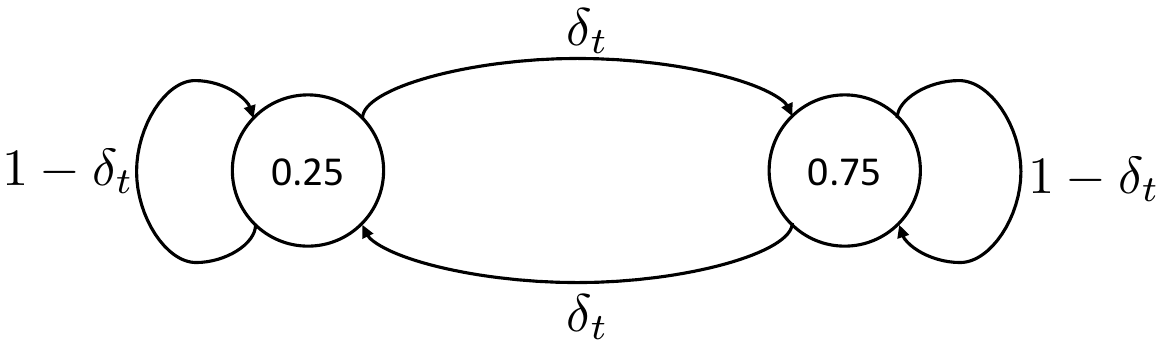}
\caption{Markov chain for the time-varying mean service rate $\mu_e(t), \forall e\in E$. At time slot $t$ and for any edge $e$, given $\mu_e(t) \in \{ 0.25, 0.75\}$, the instantaneous service capacity $\theta_e(t)$ follows Rayleigh distribution with mean $\mu_e(t)$.}
\label{fig: channel_condition}
\end{figure}

In this Section, we empirically evaluate the performance of MW-UCB and validate its throughput-optimality. 
We compare our proposed algorithm with two baseline algorithms: 
\begin{itemize}
    \item The idealized  Max-Weight (MW) policy \cite{neelybook1} with full knowledge of network statistics, which schedules the link activation vector at time slot $t$ according to  \eqref{MW_scheduling}. This represents the class of idealized policies $\Pi_s$ and serves as an \emph{unrealistic} baseline.
    \item The MW with restart UCB \cite{STAHLBUHK2019131}, which can be thought of as a special case of MW-UCB for $d = \tau$ and represents the class of admissible policies $\Pi$.
    While originally proposed for stationary network control with partial observability and unknown statistics, this  algorithm was  empirically verified as a heuristic for non-stationary settings in \cite{STAHLBUHK2019131}, and is the only algorithm in the literature that is directly applicable to our model.
\end{itemize}
\begin{figure}[H]  
\centering
\includegraphics[width=0.17\textwidth]{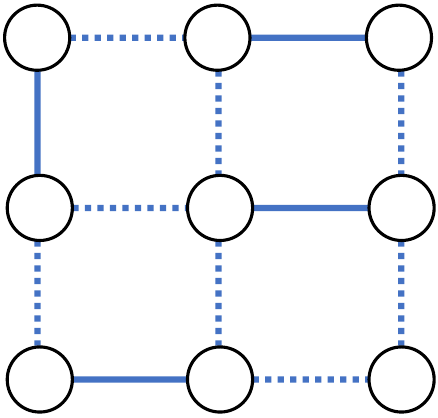}
\caption{The $3\times 3$ grid topology; an example of link schedule subject to node-exclusive interference, which forms a matching, is shown in solid lines. }
\label{fig: grid}
\end{figure}   
For both  MW-UCB and MW with restart UCB \cite{STAHLBUHK2019131}, we set the restart period to $\tau = T^{2/3}$. Sliding window size of MW-UCB is set to $d = 2 \lceil \tau^{\frac{2}{3}(1-\alpha)} \rceil + 150$.
We perform extensive testing on the $3 \times 3$ grid network with node-exclusive wireless interference constraints \cite{4967889}, as depicted in Figure \ref{fig: grid}. 
To model the non-stationary service rates, for any time slot $t$ and link $e\in E$, we consider $\mu_e(t)$ evolving over time according to the Markov chain in Figure \ref{fig: channel_condition}, whereby $\mu_e(t)$ would change its state (between $0.25$ and $0.75$) with probability $\delta_t$ which itself may vary over time. 
Then, given $\mu_e(t)$, the instantaneous service capacity $\theta_e(t) \sim  Rayleigh\big(\sqrt{\frac{2}{\pi}} \mu_e(t)\big)$ is sampled from the Rayleigh distribution  with the scale parameter that ensures  $\EE[\theta_e(t)] = \mu_e(t)$. We consider two settings of $\delta_t$, which governs the dynamics of the non-stationary service rates:
\begin{enumerate}
    \item Time-invariant $\delta_t = \frac{0.5}{T^{1/2}}$: This corresponds to the uniformly changing dynamics and was considered in the literature \cite{STAHLBUHK2019131} for simulations.
    \item Time-varying $\delta_t = \frac{0.5}{(t+1)^{1/2}}$: This  non-stationary aperiodic setting captures  more abruptly changing environments. 
\end{enumerate}
\noindent
Moreover, both of the above settings satisfy  Assumption \ref{vari_assumption} in the sense that $\EE[\gamma(t_1, t_2)] = O(|t_2 - t_1|^{1/2})$  for any $0\leq t_1 < t_2 \leq T$ \footnote{While this is $\EE[\gamma(t_1, t_2)] $,  we can strictly enforce Assumption \ref{vari_assumption}, i.e. without expectation, by deterministically simulating a feasible trajectory of $\muu(t)$'s evolution.} (see Appendix \ref{appen_observation} for the proof). We thus use $\alpha = 1/2$ in our simulations. 
\begin{figure*}[h]
\centering
\begin{subfigure}{0.315\textwidth}
\includegraphics[width=\textwidth]{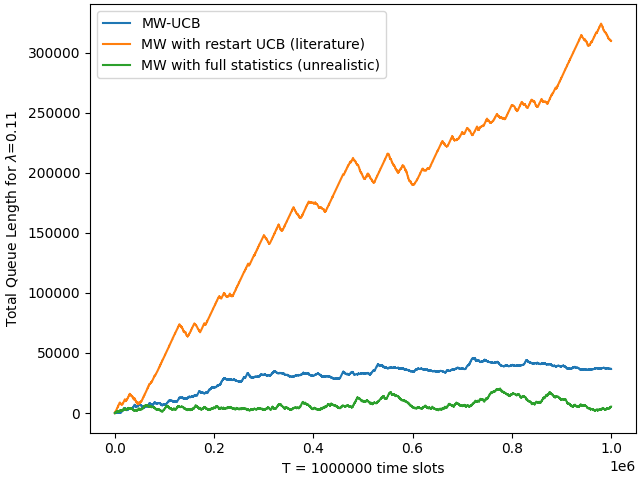}
\caption{Fixed arrival rate $\lambda = 0.11$ strictly within the stability region.}
\label{fig: uniform_delta1}
\end{subfigure}
\hspace{0.2cm}
\begin{subfigure}{0.315\textwidth}
\includegraphics[width=\textwidth]{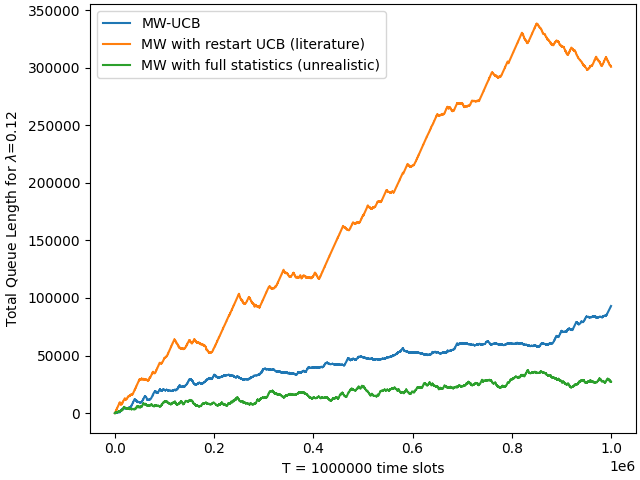}
\caption{Fixed arrival rate $\lambda = 0.12$ strictly outside the stability region.} 
\label{fig: uniform_delta2}
\end{subfigure}
\hspace{0.2cm}
\begin{subfigure}{0.315\textwidth}
\includegraphics[width=\textwidth]{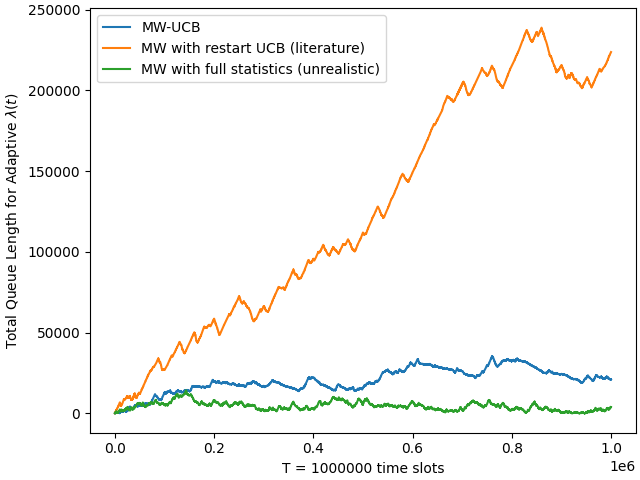}
\caption{Adaptive arrival rates $\bar{\lambda}(t)$ of highly loaded network.} 
\label{fig: adaptive_rate_uniform}
\end{subfigure}
\caption{ Total queue length over time for time-invariant   $\delta_t = \frac{0.5}{T^{1/2}}$.}
\label{fig: uniform_delta}
\end{figure*}

\subsection{Throughput-Optimality and Stability}
\begin{figure*}[ht]
\centering
\begin{subfigure}{0.315\textwidth}
\includegraphics[width=\textwidth]{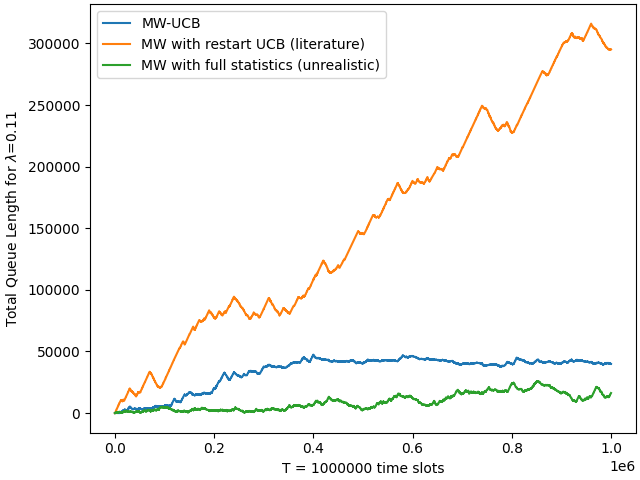}
\caption{Fixed arrival rate $\lambda = 0.11$ strictly within the stability region.}
\label{fig: non_uniform_delta1}
\end{subfigure}
\hspace{0.2cm}
\begin{subfigure}{0.315\textwidth}
\includegraphics[width=\textwidth]{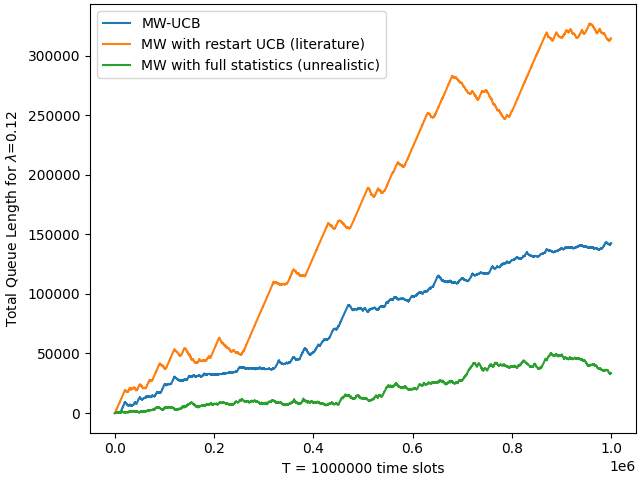}
\caption{Fixed arrival rate $\lambda = 0.12$ strictly outside  the stability region.} 
\label{fig: non_uniform_delta2}
\end{subfigure}
\hspace{0.2cm}
\begin{subfigure}{0.315\textwidth}
\includegraphics[width=\textwidth]{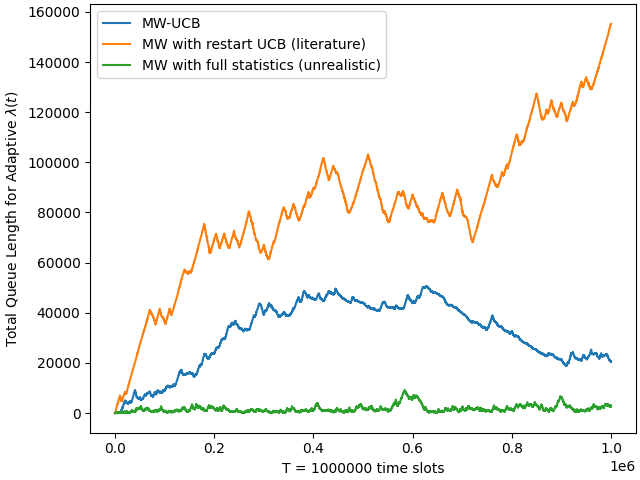}
\caption{Adaptive arrival rates $\bar{\lambda}(t)$ of highly loaded network.} 
\label{fig: adaptive_rate_non_uniform}
\end{subfigure}
\caption{Total queue length over time for time-varying   $\delta_t = \frac{0.5}{(t+1)^{1/2}}$.}
\label{fig: non_uniform_delta}
\end{figure*}

We first consider fixed arrival rates where, at time slot $t$, every link receives Poisson arrivals with the same packet generation rate  $\lambda = \lambda_e(t), \forall e\in E$. 
To demonstrate the stability properties of the algorithms, we investigate the evolution of the total queue backlog, i.e. $\sum_{e\in E} Q_e(t)$ at time $t$, for $\lambda = 0.11$ and $0.12$, which respectively represent the  regimes of moderately loaded and highly loaded network. We run the simulations for $T = 10^6$ time slots and report the results  for both settings of time-invariant $\delta_t$ and time-varying $\delta_t$   in Figure \ref{fig: uniform_delta} and Figure \ref{fig: non_uniform_delta}, respectively.


\textbf{Throughput-Optimality of MW-UCB:} The results from Figure \ref{fig: uniform_delta} and Figure \ref{fig: non_uniform_delta} demonstrate that  MW-UCB  preserves the stability property of  the idealized MW policy and thus supports the same stability region as achieved by the class of  idealized  policies  with full statistics. In particular, the total queue backlogs of both algorithms remain stable for the moderate-load regime $\lambda = 0.11$, and start to explode in the high-load regime $\lambda=0.12$.

\begin{figure*}[ht]
\centering
\begin{subfigure}{0.315\textwidth}
\includegraphics[width=\textwidth]{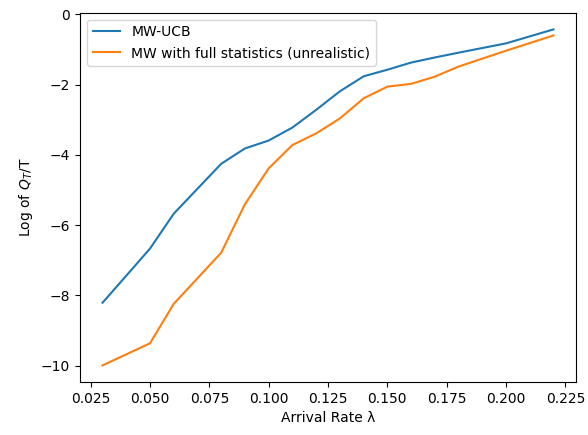}
\caption{Setting of time-invariant   $\delta_t = \frac{0.5}{T^{1/2}}$.}
\label{fig: uniform_log_QT_plot}
\end{subfigure}
\hspace{0.1cm}
\begin{subfigure}{0.307\textwidth}
\includegraphics[width=\textwidth]{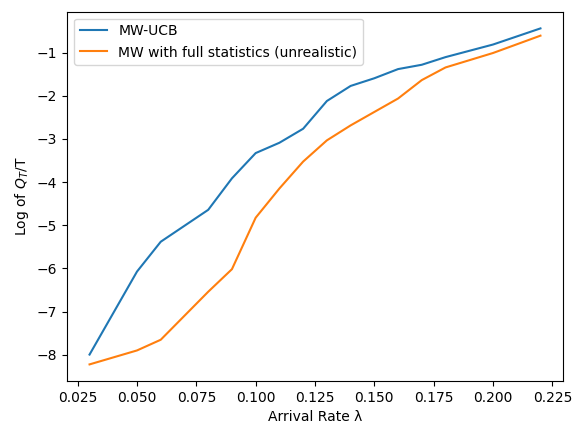}
\caption{Setting of time-varying   $\delta_t = \frac{0.5}{(t+1)^{1/2}}$.} 
\label{fig: non_uniform_log_QT_plot}
\end{subfigure}
\caption{Measuring $\log(Q_T/T)$ at $T = 1.5 \cdot 10^6$ for  $\lambda = 0.03 \to 0.22$.}
\label{fig: log_QT_plot}
\end{figure*}

\textbf{Performance Evaluation of MW-UCB:} In all experiments from Figure \ref{fig: uniform_delta} and Figure \ref{fig: non_uniform_delta}, MW-UCB  consistently outperforms MW with restart UCB. Whenever the arrival rate $\lambda$ is inside the stability region (Figure \ref{fig: uniform_delta1} and Figure \ref{fig: non_uniform_delta1}), MW-UCB can learn the channels' statistics under non-stationarity and consequently stabilize the system. Additionally, for MW-UCB and idealized MW, we gradually increase $\lambda = 0.03 \to 0.22$ and report in Figure \ref{fig: log_QT_plot} the value of $\log\big({\sum_{e\in E} Q_e(T)}/{T} \big)$ at $T = 1.5 \cdot 10^6$ to empirically measure the closeness of  ${Q_T}/{T}$ to $0$ as well as its growth outside the stability region. The result suggests that  MW-UCB  preserves the pattern of idealized MW.

\subsection{Time-Varying Arrival Rates}

Next, we provide additional simulations for time-varying arrival rates $\lmb(t)$. 
We let all links in any time slot $t$ to receive Poisson arrivals with the  same packet generation rate  $\bar{\lambda}(t) = \lambda_e(t), \forall e\in E$.
Given the mean service rates $\muu(t)$, an upper-bound on the maximum arrival rate supported by the network is given by:
\begin{align}
    \bar{\lambda}(t) \leq \min_{v\in V} \frac{1}{\sum_{e \in \Aa(v)} \mu_e(t)^{-1}},\label{time_varying_lmb1}
\end{align}

where $\Aa(v)$ is the set of links adjacent to node $v$. In our simulations, we set $\bar{\lambda}(t)$ to be exactly the right-hand side of \eqref{time_varying_lmb1} to model highly loaded network. We plot the total queue length over time, i.e. $\sum_{e\in E} Q_e(t)$ at time $t$,  for both settings of time-invariant $\delta_t$ and time-varying $\delta_t$ respectively in Figure \ref{fig: adaptive_rate_uniform} and Figure \ref{fig: adaptive_rate_non_uniform}. 
The results demonstrate that MW-UCB can well adapt to the time-varying arrival rates to achieve stability, and consistently improves over MW with restart UCB.

\section{Conclusion}
\label{sec:conclusion}

In this paper, we present MW-UCB as a novel joint learning and scheduling algorithm for non-stationary wireless network control under partial observability and non-stationary dynamics. 
Our algorithmic development is based on the Max-Weight policy for network control and sliding-window UCB for learning uncertain and time-varying channel statistics. 
We propose a new notion of stability for non-stationary networks and prove that the MW-UCB algorithm  achieves a stability region that is arbitrarily close to the true stability region. Extensive simulations on both uniformly changing and abruptly changing dynamics confirm the throughput-optimality and the favorable performance of the algorithm.   We believe that our analytical framework can be extended to study  stability properties of algorithms for non-stationary network control under stringent requirements of emerging large-scale wireless networks.

\section*{Acknowledgment}
This work was supported by ONR grant N00014-20-1-2119.


\bibliographystyle{ACM-Reference-Format}
\bibliography{sample-base}


\newpage

\appendix

\section*{Appendix}

\section{Proof of Lemma \ref{lma:approx_region1}}
\label{appen_approx_region1}

First, we prove that $ \Lmb_s(\beta_1) \subseteq  \Lmb_s(\beta_2)$ for any $0\leq \beta_1 \leq \beta_2 \leq 1$. Take any $\{\lmb(t) \}_{t\geq 0} \in  \Lmb_s(\beta_1)$. Then there exists some $\pi \in \Pi_s$  such that $Q_T^\pi = O\big(T^{\beta_1} \big)$. Since $\beta_1 \leq \beta_2$, this implies  $Q_T^\pi = O\big(T^{\beta_2} \big)$ and thus $\{\lmb(t) \}_{t\geq 0} \in  \Lmb_s(\beta_2)$. Therefore, we have $ \Lmb_s(\beta_1) \subseteq  \Lmb_s(\beta_2)$.

Second, we prove that $ \Lmb_s(\beta) \subseteq \Lmb_s $ for any $\beta \in [0, 1)$. Take any $\{\lmb(t) \}_{t\geq 0} \in  \Lmb_s(\beta)$. Then there exists some $\pi \in \Pi_s$  such that $Q_T^\pi = O\big(T^{\beta} \big)$. Since $\beta <1 $, this implies  $Q_T^\pi = o\big(T \big)$ and thus $\{\lmb(t) \}_{t\geq 0} \in  \Lmb_s$. Therefore, we have $ \Lmb_s(\beta) \subseteq  \Lmb_s$.

Finally, we prove that $ \Lmb_s \subseteq \Lmb_s(1)$. Take any $\{\lmb(t) \}_{t\geq 0} \in  \Lmb_s$. Then there exists some $\pi \in \Pi_s$  such that $Q_T^\pi = o\big(T \big)$, which implies  $Q_T^\pi = O\big(T \big)$ and thus $\{\lmb(t) \}_{t\geq 0} \in  \Lmb_s(1)$. Therefore, we have $ \Lmb_s \subseteq  \Lmb_s(1)$.

\section{Proof of Theorem \ref{thrm:equivalence}}
\label{appen_equivalence}


Since $\lmb \in \bar{\Lmb}_s$, by definition there exists some $\pi^* \in \Pi_s$ that satisfies:  
\begin{align}
    \lambda_e = \lim_{T\to \infty} \frac{1}{T} \sum_{t=0}^{T-1} a_e(t) \leq  \lim_{T\to \infty} \frac{1}{T} \sum_{t=0}^{T-1} b^{\pi^*}_e(t), \forall e \in E. \label{equivalence_proof1}
\end{align}
Let $\xX^1, \xX^2, ..., \xX^{|\Mm|}$ be all the admissible link activations in $\Mm$. For any time slot $t > 0$, we consider the empirical counter and distribution $\forall i \in [1, |\Mm|]$:
\begin{align}
    n_i(t) &=  \sum_{s=0}^{t-1} \mathbbm{1}_{\{\xX^{\pi^*}(s) = \xX^i \}}, \label{equivalence_proof2}\\
    p_i(t) &= \frac{ n_i(t) }{t}, \label{equivalence_proof3}
\end{align}
where $ n_i(t)$ and $p_i(t)$ respectively represent the number of times and the time fraction   that the link activation vector $\xX^i$ has been chosen by the policy $\pi^*$ by time $t$. Additionally, we consider $\forall i \in [1, |\Mm|]$:
\begin{align}
    S_{i, e}(t) = \begin{cases}
        0, \text{ if $x^i_e = 0$}\\
        \frac{  \sum_{s=0}^{t-1} \theta_e(s) \mathbbm{1}_{\{\xX^{\pi^*}(s) = \xX^i \}} }{ n_i(t) }, \text{ otherwise}
    \end{cases}. \label{equivalence_proof4}
\end{align}
Let $\pP(t) = ( p_i(t) )_{i\in [1, |\Mm|]}$ be the vector of all such empirical distributions. Also note that $\sum_{i=1}^{|\Mm|} p_i(t) = 1$ for any $t > 0$. Since $\pP(t) \in [0, 1]^{ |\Mm |}$, by the Bolzano–Weierstrass Theorem, there exists a convergent subsequence $\{ \pP(t_k) \}_{k = 1, 2,...}$ where $\lim_{k\to \infty} t_k = \infty$. Define the limit of this convergent subsequence by:
\begin{align}
    \pP^* = \lim_{k\to \infty} \pP(t_k). \label{equivalence_proof5}
\end{align}
Now, we note that for any $i \in [1, |\Mm|]$ such that $p^*_i > 0$ (which also implies $\lim_{k\to \infty} n_i(t_k) = \infty$), by SLLN, we have:
\begin{align}
    \lim_{k\to \infty} S_{i, e}(t_k) =  \mathbbm{1}_{\{ x^i_e = 1 \}} \mu_e, \forall e \in E.  \label{equivalence_proof6}
\end{align}
With the convention that $0/0 = 0$, we obtain from \eqref{equivalence_proof1} that $\forall e\in E$:
\begin{align}
\nonumber
     \lambda_e &\leq  \lim_{k \to \infty} \frac{1}{t_k} \sum_{s=0}^{t_k-1} b^{\pi^*}_e(s) \\
     \nonumber
     &= \lim_{k \to \infty} \sum_{i=1}^{|\Mm|}  p_i(t_k) \cdot  S_{i, e}(t_k) \\
     &\overset{\eqref{equivalence_proof5} + \eqref{equivalence_proof6} }{=}  \mu_e \sum_{i=1}^{|\Mm|} p^*_i \cdot  \mathbbm{1}_{\{ x^i_e = 1 \}} \label{equivalence_proof7}
\end{align}
Consider a stationary randomized policy $\pi_r$ that at any time $t$ activates the link schedule $\xX^{\pi_r}(t) = \xX^i$ with probability $p^*_i$. Next, we consider the Max-Weight (MW) policy with the full knowledge of the statistics (i.e. \textsc{MW} is in $\Pi_s$) that at any time slot $t$, schedules the link according to:
\begin{align}
    \xX^{\textsc{MW}}(t) =   \argmaxE_{\xX \in \Mm} \big\{ \sum_{e\in E} Q^\textsc{MW}_e(t) \mu_e x_e \big\}, \label{MW_proof1}
\end{align}
and proceeds to show that $Q_T^{ \textsc{MW} } = O\big(T^{\frac{1}{2}} \big)$. 
For brevity, we use $\pi_M$ to denote the MW policy.
Under the MW policy, we consider the  quadratic Lyapunov function of the queue lengths $\QQ^{\pi_M}(t)$ as:
\begin{align}
\label{lyapunov_potential_MW}
      \Phi^{\pi_M}(\QQ^{\pi_M}(t)) = \QQ^{\pi_M}(t)^T \QQ^{\pi_M}(t) = \sum_{e\in E} Q^{\pi_M}_e(t)^2.
\end{align}
We  consider the $1$-step Lyapunov drift  conditioned on the queue length as follows: 
\begin{align}
\label{lyapunov_drift1_MW}
    \Delta^{\pi_M} (t) = \EE\big[\Phi^{\pi_M}(\QQ^{\pi_M}(t+1)) - \Phi^{\pi_M}(\QQ^{\pi_M}(t)) \big|   \QQ^{\pi_M}(t) \big].
\end{align}
From the queue process \eqref{queue_dynamics1}, we first obtain that $\forall e \in E, t\geq 0$:
\begin{align*}
   Q^{\pi_M}_e(t+1)^2 &\leq   \big(Q^{\pi_M}_e(t) + {a}_e(t) - b^{\pi_M}_e(t) \big)^2 \\
   &= Q^{\pi_M}_e(t)^2  + ( {a}_e(t) - b^{\pi_M}_e(t))^2 \\
   &\quad + 2 Q^{\pi_M}_e(t) ( {a}_e(t) - b^{\pi_M}_e(t)) \\
   &\leq Q^{\pi_M}_e(t)^2 + (A_{max} + \mu_{max})^2   \\
   &\quad + 2 Q^{\pi_M}_e(\tau_j) ( {a}_e(t) - b^{\pi_M}_e(t)),
\end{align*}
where in the last line we use  $0\leq a_e(t) \leq A_{max}$, $ 0\leq b^{\pi_M}_e(t) \leq \theta_e(t) \leq \mu_{max}$. 
Summing the above over all $e\in E$ and taking the expectation conditioned on   $\QQ^{\pi_M}(t)$, we obtain that: 
\begin{align*}
    \Delta^{\pi_M} (t) &\leq   |E|  (A_{max} + \mu_{max})^2+ 2 \sum_{e\in E} Q^{\pi_M}_e(t) \lambda_e(t) \\
    &\quad - 2 \sum_{e\in E} Q^{\pi_M}_e(t) \mu_e \EE\big[x^{\pi_M}_e(t) \big| \QQ^{\pi_M}(t) \big] \\
    &\overset{\eqref{MW_proof1}}{\leq}  |E|  (A_{max} + \mu_{max})^2+ 2 \sum_{e\in E} Q^{\pi_M}_e(t) \lambda_e(t) \\
    &\quad - 2 \sum_{e\in E} Q^{\pi_M}_e(t) \mu_e \EE\big[x^{\pi_r}_e(t) \big| \QQ^{\pi_M}(t) \big] \\
    &=  |E|  (A_{max} + \mu_{max})^2+ 2 \sum_{e\in E} Q^{\pi_M}_e(t) \lambda_e(t) \\
    &\quad - 2 \sum_{e\in E} Q^{\pi_M}_e(t) \mu_e  \sum_{i=1}^{|\Mm|} p^*_i \cdot  \mathbbm{1}_{\{ x^i_e = 1 \}}\\
    &\overset{\eqref{equivalence_proof7}}{\leq}  |E|  (A_{max} + \mu_{max})^2.
\end{align*}
Taking expectation of both sides  with respect to $\QQ^{\pi_M}(t)$, we have:
\begin{align*}
\EE\big[   \Phi^{\pi_M}(\QQ^{\pi_M}(t+1))\big] - \EE\big[   \Phi^{\pi_M}(\QQ^{\pi_M}(t))\big] \leq |E|  (A_{max} + \mu_{max})^2.
\end{align*}
Telescoping over $t = 0 \to T-1$ and noting that $ \Phi^{\pi_M}(\QQ^{\pi_M}(0)) = 0$, we get:
\begin{align}
     \EE\big[   \Phi^{\pi_M}(\QQ^{\pi_M}(T))\big] \leq T |E|  (A_{max} + \mu_{max})^2. \label{equivalence_proof8}
\end{align}
Following the same argument as the proof of Lemma \ref{lma:queue_to_potential}, we can similarly obtain that:
\begin{align*}
\nonumber
    Q_T^{\textsc{MW}} = \EE\big[\sum_{e\in E} Q^{\pi_M}_e(T) \big] &\leq \sqrt{|E| \cdot \EE\big[ \Phi^{\pi_M}(\QQ^{\pi_M}(T)) \big]} \\
    &\overset{ \eqref{equivalence_proof8} }{\leq} T^{\frac{1}{2}} |E| (A_{max} + \mu_{max}),
\end{align*}
which  implies $Q_T^{ \textsc{MW} } = O\big(T^{\frac{1}{2}} \big)$. Since $\pi_M \in \Pi_s$, we thus have $\lmb \in \Lmb_s(\frac{1}{2})$. Since by our Definition  \ref{def:throughput}, a throughput-optimal policy $\pi$, given $\beta \in [0, 1)$,  achieves the stability region $\Lmb_s(\beta)$, the network is  mean rate stable under $\pi$ for any $\lmb\in \bar{\Lmb}_s$.

\section{Proof of Lemma \ref{lma:shedding1}}
\label{appen_shedding}

Since $\{\lmb(t) \}_{t\geq 0} \in  \Lmb_s(\beta)$, by definition there exists some policy $\pi^* \in \Pi_s$ such that:
\begin{align*}
    Q_T^{\pi^*} = O(T^\beta).
\end{align*}
For any $\tau_j$, we consider  $\forall e \in E$:
\begin{align}
    r_e(\tau_j) =  \begin{cases} 0,  \text{ if } \sum_{t=\tau_j}^{\tau_{j+1}-1} \lambda_e(t) = 0 \\ 
    \bigg( \frac{(1-\varepsilon) (\sum_{t=\tau_j}^{\tau_{j+1}-1} \lambda_e(t) -\EE[Q_e^{\pi^*}(\tau_{j+1})])}{ \sum_{t=\tau_j}^{\tau_{j+1}-1} \lambda_e(t) } \bigg)^+,\text{ otherwise } 
    \end{cases}
     \label{shedding_inter1}
\end{align}
and sheds the traffics such that for any $t \in [\tau_j, \tau_{j+1})$,
\begin{align}
    \tilde{a}_e(t) = r_e(\tau_j) a_e(t), \forall e\in E. \label{shedding_inter2}
\end{align}
We now proceed to show that  this shedding scheme guarantees $\{\tilde{\lmb}(t) \}_{t\geq 0} \in  \Cc(\tau, \varepsilon)$ and \eqref{shedding_cost}. 

From the queue dynamics \eqref{queue_dynamics1}, we have $\forall e \in E$:
\begin{align*}
    Q_e^{\pi^*}(\tau_{j+1}) &= \big(Q^{\pi^*}_e(\tau_{j+1}-1) + a_e(\tau_{j+1}-1) - b^{\pi^*}_e(\tau_{j+1}-1) \big)^+\\
    &\geq Q^{\pi^*}_e(\tau_{j+1}-1) + a_e(\tau_{j+1}-1) - b^{\pi^*}_e(\tau_{j+1}-1).
\end{align*}
By repeating the above argument, we get $\forall e \in E$:
\begin{align*}
    Q_e^{\pi^*}(\tau_{j+1}) &\geq Q^{\pi^*}_e(\tau_{j}) + \sum_{t= \tau_j}^{\tau_{j+1}-1} a_e(t) -\sum_{t= \tau_j}^{\tau_{j+1}-1}  b^{\pi^*}_e(t)\\ 
     &\geq \sum_{t= \tau_j}^{\tau_{j+1}-1} a_e(t) -\sum_{t= \tau_j}^{\tau_{j+1}-1}  b^{\pi^*}_e(t).
\end{align*}
Taking expectation of the above, we obtain that:
\begin{align}
\nonumber 
    \sum_{t=\tau_j}^{\tau_{j+1}-1} \lambda_e(t) - \EE[ Q_e^{\pi^*}(\tau_{j+1})] \leq   \sum_{t= \tau_j}^{\tau_{j+1}-1}  \EE[b^{\pi^*}_e(t)], \quad \forall e \in E. \\
   \therefore \bigg( \sum_{t=\tau_j}^{\tau_{j+1}-1} \lambda_e(t) - \EE[ Q_e^{\pi^*}(\tau_{j+1})]  \bigg)^+ \leq   \sum_{t= \tau_j}^{\tau_{j+1}-1}  \EE[b^{\pi^*}_e(t)], \quad \forall e \in E. \label{shedding_inter3}
\end{align}
Taking expectation of  \eqref{shedding_inter2} and summing over $t = \tau_j \to \tau_{j+1}-1$, we obtain in view of  \eqref{shedding_inter1} and \eqref{shedding_inter3} that $\forall e\in E$:
\begin{align*}
     \sum_{t=\tau_j}^{\tau_{j+1}-1} \tilde{\lambda}_e(t) \overset{\eqref{shedding_inter2}}{=} r_e(\tau_j) \big( \sum_{t=\tau_j}^{\tau_{j+1}-1} \lambda_e(t) \big) \overset{\eqref{shedding_inter1}+\eqref{shedding_inter3}}{\leq}  (1-\varepsilon)  \sum_{t= \tau_j}^{\tau_{j+1}-1}  \EE[b^{\pi^*}_e(t)],
\end{align*}
which verifies $\{\tilde{\lmb}(t) \}_{t\geq 0} \in  \Cc(\tau, \varepsilon)$.

To prove \eqref{shedding_cost}, we  note that from \eqref{XT}, $\EE[ X_T ] = \sum_{t=0}^{T-1} \sum_{e\in E} \lambda_e(t)-\sum_{t=0}^{T-1} \sum_{e\in E} \tilde{\lambda}_e(t)$, 
and first proceeds to upper-bound $\sum_{t=\tau_j}^{\tau_{j+1}-1}\lambda_e(t)-\sum_{t=\tau_j}^{\tau_{j+1}-1} \tilde{\lambda}_e(t)$ by considering two cases as follows.\\
\textbf{Case 1:} $\sum_{t=\tau_j}^{\tau_{j+1}-1} \lambda_e(t) \leq \EE[Q_e^{\pi^*}(\tau_{j+1})]$. From \eqref{shedding_inter1}, we have $r_e(\tau_j) = 0$ and thus (from  \eqref{shedding_inter2})  $\tilde{\lambda}_e(t) = \EE[\tilde{a}_e(t)]= 0$. Then we obtain that:
\begin{align*}
    \sum_{t=\tau_j}^{\tau_{j+1}-1}\lambda_e(t)-\sum_{t=\tau_j}^{\tau_{j+1}-1} \tilde{\lambda}_e(t) = \sum_{t=\tau_j}^{\tau_{j+1}-1}\lambda_e(t) \leq \EE[Q_e^{\pi^*}(\tau_{j+1})]. 
\end{align*}
\textbf{Case 2:} $\sum_{t=\tau_j}^{\tau_{j+1}-1} \lambda_e(t) > \EE[Q_e^{\pi^*}(\tau_{j+1})] \geq 0$. We have:
\begin{align*}
     \sum_{t=\tau_j}^{\tau_{j+1}-1}\lambda_e(t)-\sum_{t=\tau_j}^{\tau_{j+1}-1} \tilde{\lambda}_e(t) &\overset{  \eqref{shedding_inter2} }{=} (1-r_e(\tau_j))  \big( \sum_{t=\tau_j}^{\tau_{j+1}-1} \lambda_e(t) \big)\\
     &\overset{  \eqref{shedding_inter1} }{=} \varepsilon  \big( \sum_{t=\tau_j}^{\tau_{j+1}-1} \lambda_e(t) \big) + (1-\varepsilon) \EE[Q_e^{\pi^*}(\tau_{j+1})]\\
     &\leq \varepsilon  \tau A_{max} +  \EE[Q_e^{\pi^*}(\tau_{j+1})] 
\end{align*}
Thus, in any case, we have:
\begin{align*}
  \sum_{t=\tau_j}^{\tau_{j+1}-1}\lambda_e(t)-\sum_{t=\tau_j}^{\tau_{j+1}-1} \tilde{\lambda}_e(t) \leq   \varepsilon  \tau A_{max} +  \EE[Q_e^{\pi^*}(\tau_{j+1})]. 
\end{align*}
Summing the above over all $e\in E$ and $j = 0 \to K-1$, we obtain:
\begin{align}
\nonumber
     \EE[ X_{\tau_K} ]  &= \sum_{t=0}^{\tau_K-1} \sum_{e\in E} \lambda_e(t)-\sum_{t=0}^{\tau_K-1} \sum_{e\in E} \tilde{\lambda}_e(t) \\
     \nonumber
     &\leq  \varepsilon  \tau K A_{max} + \sum_{j=0}^{K-1} \EE[ \sum_{e\in E} Q_e^{\pi^*}(\tau_{j+1})  ] \\
     \nonumber
     &=  \varepsilon  \tau K A_{max} + \sum_{j=1}^{K} Q^{\pi^*}_{\tau_j}\\
     \nonumber
     &= O\big( \varepsilon  \tau K  + \sum_{j=1}^{K} \tau_j^\beta  \big)\\
     \nonumber 
     &= O\big( \varepsilon  \tau K  + \tau_K^\beta+ \sum_{j=1}^{K-1} \frac{1}{\tau} \int_{\tau_j}^{\tau_{j+1}} x^\beta dx  \big)\\
     \nonumber
     &= O\big( \varepsilon  \tau K  + \tau_K^\beta+ \frac{1}{\tau} \int_{\tau_1}^{\tau_{K}} x^\beta dx  \big)\\
     \nonumber
     &= O\big( \varepsilon  \tau K  + \tau_K^\beta+ \frac{1}{\tau} \int_{\tau_1}^{\tau_{K}} x^\beta dx  \big)\\
     \nonumber
     &= O\big( \varepsilon  \tau_K  + \tau_K^\beta+ \frac{\tau_K^{\beta+1}}{\tau}   \big)\\
     &= O\big(   T^{\beta+1}\tau^{-1}  + \varepsilon  T  \big).  \label{shedding_inter4}
\end{align}
Finally, we have:
\begin{align*}
     \EE[ X_{T} ] &=  \EE[ X_{\tau_K} ] + \sum_{t=\tau_K}^{T-1}  \sum_{e\in E} \lambda_e(t)-\sum_{t=\tau_K}^{T-1} \sum_{e\in E} \tilde{\lambda}_e(t)\\
     &\leq \EE[ X_{\tau_K} ] + \tau |E| A_{max} \\
     &\overset{ \eqref{shedding_inter4} }{=}  O\big(   T^{\beta+1}\tau^{-1}  + \varepsilon  T +\tau \big).
\end{align*}

\section{Guarantees of the imaginary system's queue process $\{ \WQQ(t)\}_{t\geq 0}$}
\label{appen_imaginary_system}

Recall from Section \ref{sec:theoretical_results} that: 
\begin{itemize}
    \item When MW-UCB is applied to the original system, it produces the sequence of decisions $\{\xX^{\textsc{MW-UCB}}(t)\}_{t\geq 0}$ and thus effective service rate $\{\bB^{\textsc{MW-UCB}}(t)\}_{t\geq 0}$. The queueing dynamics of the original system evolves as $\{ \QQ^{\textsc{MW-UCB}}(t) \}_{t\geq 0}$ via:
    \begin{align*}
        Q^{\textsc{MW-UCB}}_e(t+1) =  \big(Q^{\textsc{MW-UCB}}_e(t) + a_e(t) - b^{\textsc{MW-UCB}}_e(t) \big)^+,  \forall e\in E. 
    \end{align*}
    \item The "imaginary" queue lengths $\WQQ(t)$ evolve as  the sequence of decisions $\{\xX^{\textsc{MW-UCB}}(t)\}_{t\geq 0}$ is applied to the imaginary system, i.e.
    \begin{align*}
        \tq_e(t+1) =  \big(\tq_e(t) + \tilde{a}_e(t) - b^{\textsc{MW-UCB}}_e(t) \big)^+,  \forall e\in E.
    \end{align*}
\end{itemize}
In the proofs of this Appendix \ref{appen_imaginary_system}, we refer to $\{\xX^{\textsc{MW-UCB}}(t)\}_{t\geq 0}$, $\{\bB^{\textsc{MW-UCB}}(t)\}_{t\geq 0}$ and $\{ \QQ^{\textsc{MW-UCB}}(t) \}_{t\geq 0}$ respectively as $\{\xX(t)\}_{t\geq 0}$, $\{\bB(t)\}_{t\geq 0}$ and $\{ \QQ(t) \}_{t\geq 0}$ for brevity. Consequently, the queueing dynamics of the original system and imaginary system can be respectively expressed as:
\begin{align}
    Q_e(t+1) =  \big(Q_e(t) + a_e(t) - b_e(t) \big)^+,  \forall e\in E \label{original_sys_queue}\\
    \tq_e(t+1) =  \big(\tq_e(t) + \tilde{a}_e(t) - b_e(t) \big)^+,  \forall e\in E. \label{imaginary_sys_queue}
\end{align}


\subsection{Proof of Lemma \ref{lma:relate_to_imag}}
\label{appen_relate_to_imag}

From Lemma \ref{lma:general_queue_properties}, we have $\forall e\in E$: 
\begin{align*}
    Q_e(T) \leq \tq_e(T) + \sum_{q= 0}^{T-1} (a_e(q) - \tilde{a}_e(q))
\end{align*}
Summing up the above over all $e\in E$ and taking expectation, we conclude that: 
\begin{align*}
    &\EE[ \sum_{e\in E} Q_e(T) ] \leq \EE[ \sum_{q= 0}^{T-1} (a_e(q) - \tilde{a}_e(q)) ] + \EE\big[ \sum_{e\in E} \widetilde{Q}_e(T)\big]  \\
     \therefore &Q_T^{\textsc{MW-UCB}} \leq \EE[X_T] + \EE\big[ \sum_{e\in E} \widetilde{Q}_e(T)\big],
\end{align*}
where the last line follows from the definitions of $Q_T$ and $X_T$.

\subsection{Proof of Lemma \ref{lma:imag_queue_bound}}
\label{appen_imag_queue_bound}

Since $\{\tilde{\lmb}(t) \}_{t\geq 0} \in  \Cc(\tau, \varepsilon)$, by definition there exists some $\pi_0 \in \Pi_s $ such that for any $\tau_j \in \{\tau_0, \tau_1,..., \tau_{K-1}\}$:
\begin{align}
\sum_{t=\tau_j}^{\tau_{j+1}-1} \tilde{\lambda}_e(t) \leq (1-\varepsilon) \sum_{t=\tau_j}^{\tau_{j+1}-1} \EE[b^{\pi_0}_e(t)], \text{ } \forall e \in E. \label{window_stationary1}   
\end{align}

We consider the quadratic Lyapunov function of the queue lengths $\WQQ(t)$ of the imaginary system under MW-UCB as:
\begin{align}
\label{lyapunov_potential}
      \Phi(\WQQ(t)) = \WQQ(t)^T \WQQ(t) = \sum_{e\in E} \tq_e(t)^2.
\end{align}
We  consider the $\tau$-step Lyapunov drift of $\Phi(.)$ conditioned on the queue lengths of both the original system and the imaginary system as follows: 
\begin{align}
\label{lyapunov_drift1}
    \Delta (\tau_j) = \EE\big[\Phi(\WQQ(\tau_{j+1})) - \Phi(\WQQ(\tau_{j})) \big|  \WQQ(\tau_{j}), \QQ(\tau_{j}) \big],
\end{align}
where we recall that $\tau_j = j\tau$. 
From    Lemma \ref{lma:drift_bound1} (in Appendix \ref{append_drift_bound1}),    the drift can be upper-bounded  by:

\begin{align}
\nonumber
     \Delta (\tau_j) &\leq B_1 \tau (\tau+1) + 2 \sum_{t= \tau_j}^{\tau_{j+1}-1} \WQQ(\tau_j)^T \wlmb(t) \\
     &\quad  - 2 \sum_{t= \tau_j}^{\tau_{j+1}-1} \EE[ \WQQ(\tau_j)^T \bB(t)\big| \WQQ(\tau_{j}), \QQ(\tau_{j}) ], \label{drfit1_restated}
\end{align}
where $B_1 = \frac{1}{2} |E|  (A_{max} + \mu_{max})^2$. Now,  we consider the normalized queue lengths of both the original and imaginary systems as: 
\begin{align}
    \tilde{w}_e(\tau_j) = \frac{\tq_e(\tau_j)}{\| \WQQ(\tau_j) \|_\infty}, \forall e\in E, \label{normalized_imaginary} \\
    w_e(\tau_j) = \frac{Q_e(\tau_j)}{\| \QQ(\tau_j) \|_\infty}, \forall e\in E,\label{normalized_original}
\end{align}
with the convention that $0/0=0$. 
Note that \eqref{normalized_original} is the same weight updating rule as Line \ref{reset_weight1} of MW-UCB  (Algorithm \ref{alg:MW_UCB}). Then we consider the following two regrets which respectively use \eqref{normalized_imaginary} and \eqref{normalized_original} in their weight instantiations:
\begin{align}
\nonumber
   \bar{\Rr}(\tau_j) &=  \sum_{t= \tau_j}^{\tau_{j+1}-1} \max_{\xX'\in \Mm} \EE\big[\sum_{e\in E}  \tilde{w}_e(\tau_j) x'_e \theta_e(t)  \big|\WQQ(\tau_{j}), \QQ(\tau_{j}) \big] \\
   &\quad - \sum_{t= \tau_j}^{\tau_{j+1}-1} \EE\big[\sum_{e\in E}  \tilde{w}_e(\tau_j) x_e(t) \theta_e(t)  \big|\WQQ(\tau_{j}), \QQ(\tau_{j}) \big] \label{imaginary_regret1} \\
   \nonumber
   &= \frac{1}{\| \WQQ(\tau_j) \|_\infty} \sum_{t= \tau_j}^{\tau_{j+1}-1} \max_{\xX'\in \Mm} \EE\big[\sum_{e\in E}  \tq_e(\tau_j) x'_e \theta_e(t)  \big|\WQQ(\tau_{j}), \QQ(\tau_{j}) \big] \\
   &\quad - \frac{1}{\| \WQQ(\tau_j) \|_\infty} \sum_{t= \tau_j}^{\tau_{j+1}-1} \EE\big[ \WQQ(\tau_j)^T \bB(t)  \big| \WQQ(\tau_{j}), \QQ(\tau_{j}) \big],  \label{imaginary_regret2}
\end{align}
and (recalling from \eqref{original_regret0}),
\begin{align}
\nonumber
   \Rr(\tau_j) &=  \sum_{t= \tau_j}^{\tau_{j+1}-1} \max_{\xX'\in \Mm} \EE\big[\sum_{e\in E} w_e(\tau_j) x'_e \theta_e(t)  \big|\WQQ(\tau_{j}), \QQ(\tau_{j}) \big] \\
   &\quad - \sum_{t= \tau_j}^{\tau_{j+1}-1} \EE\big[\sum_{e\in E}  w_e(\tau_j) x_e(t) \theta_e(t)  \big|\WQQ(\tau_{j}), \QQ(\tau_{j}) \big] \label{original_regret1}\\
   \nonumber 
   &= \frac{1}{\| \QQ(\tau_j) \|_\infty} \sum_{t= \tau_j}^{\tau_{j+1}-1} \max_{\xX'\in \Mm} \EE\big[\sum_{e\in E}  Q_e(\tau_j) x'_e \theta_e(t)  \big|\WQQ(\tau_{j}), \QQ(\tau_{j}) \big] \\
   &\quad - \frac{1}{\| \QQ(\tau_j) \|_\infty} \sum_{t= \tau_j}^{\tau_{j+1}-1} \EE\big[ \QQ(\tau_j)^T \bB(t)  \big|\WQQ(\tau_{j}), \QQ(\tau_{j}) \big]. \label{original_regret2}
\end{align}
From Lemma \ref{lma:relate_imaginary_regret1} (in Appendix \ref{append_drift_bound1}), we   further relate \eqref{drfit1_restated} to the regret $  \bar{\Rr}(\tau_j)$ as follows:
\begin{align}
\nonumber 
  &\sum_{t= \tau_j}^{\tau_{j+1}-1} \EE\big[ \WQQ(\tau_j)^T \bB(t)  \big| \WQQ(\tau_{j}), \QQ(\tau_{j}) \big] \\
  \nonumber 
  &\geq  (1-\varepsilon) \sum_{t= \tau_j}^{\tau_{j+1}-1} \EE\big[ \WQQ(\tau_j)^T \bB^{\pi_0}(t)  \big| \WQQ(\tau_{j}), \QQ(\tau_{j}) \big] \\
  \nonumber
  &\quad  + \varepsilon  \tau \| \WQQ(\tau_j) \|_\infty \mu_{min} - \| \WQQ(\tau_j) \|_\infty   \bar{\Rr}(\tau_j)\\
   \nonumber 
  &=  \sum_{e\in E}  \tq_e(\tau_j) \big[ (1-\varepsilon) \sum_{t= \tau_j}^{\tau_{j+1}-1} \EE[  b_e^{\pi_0}(t) ]  \big] \\
  \nonumber
  &\quad  + \varepsilon  \tau \| \WQQ(\tau_j) \|_\infty \mu_{min} - \| \WQQ(\tau_j) \|_\infty   \bar{\Rr}(\tau_j)\\
  \nonumber 
  &\overset{\eqref{window_stationary1}}{\geq}  \sum_{e\in E}  \tq_e(\tau_j) \big[  \sum_{t= \tau_j}^{\tau_{j+1}-1} \tilde{\lambda}_e(t) \big]  \\
  &\quad + \varepsilon  \tau \| \WQQ(\tau_j) \|_\infty \mu_{min} - \| \WQQ(\tau_j) \|_\infty   \bar{\Rr}(\tau_j) \label{relate_regret1} \\
  &= \sum_{t= \tau_j}^{\tau_{j+1}-1} \WQQ(\tau_j)^T \wlmb(t)   + \varepsilon  \tau \| \WQQ(\tau_j) \|_\infty \mu_{min} - \| \WQQ(\tau_j) \|_\infty   \bar{\Rr}(\tau_j) \label{relate_regret2}
\end{align}
where for \eqref{relate_regret1} we recall that $\pi_0 \in \Pi_s$ is the idealized policy that satisfies \eqref{window_stationary1}. Combining \eqref{relate_regret2} and \eqref{drfit1_restated}, we have:
\begin{align}
 \Delta (\tau_j) &\leq B_1 \tau (\tau+1)  - 2   \varepsilon  \tau \| \WQQ(\tau_j) \|_\infty \mu_{min} + 2 \| \WQQ(\tau_j) \|_\infty   \bar{\Rr}(\tau_j).   \label{inter_drift1}  
\end{align}
Now,  we have from Lemma \ref{lma:relate_two_regrets} (in Appendix \ref{append_drift_bound1}) that:
\begin{align}
    \| \WQQ(\tau_j) \|_\infty   \bar{\Rr}(\tau_j) \leq \| \QQ(\tau_j) \|_\infty   \Rr(\tau_j) + \tau \mu_{max} \EE[X_{\tau_j}]  .    \label{relate_two_regrets_restated}
\end{align}
and from Lemma \ref{lma:general_queue_properties} (in Appendix \ref{append_drift_bound1}) that:
\begin{align}
 \EE\big[ \|\WQQ(\tau_{j})\|_\infty    \big|\WQQ(\tau_{j}), \QQ(\tau_{j}) \big]    \geq \EE\big[ \|\QQ(\tau_{j})\|_\infty  \big|\WQQ(\tau_{j}), \QQ(\tau_{j}) \big]   - \EE[X_{\tau_{j}}]    \label{relate_vq4_restated}
\end{align}
Plugging \eqref{relate_two_regrets_restated} and \eqref{relate_vq4_restated} into \eqref{inter_drift1}, we obtain that:
\begin{align}
\nonumber
    \Delta (\tau_j) &\leq B_1 \tau (\tau+1) +  2 (\mu_{max} + \varepsilon \mu_{min})\tau \EE[X_{\tau_{j}}]   \\
    &\quad - 2   \varepsilon  \tau \| \QQ(\tau_j) \|_\infty \mu_{min} + 2 \| \QQ(\tau_j) \|_\infty   \Rr(\tau_j). \label{inter_drift2}
\end{align}
Recall from Section \ref{sec:alg_develop} that MW-UCB, during every time frame $[\tau_j, \tau_{j+1})$, fixes the queue length in the original system to $\QQ(\tau_j)$ (and thus the normalized weights $\{ w_e(\tau_j)\}_{e\in E}$), and  adopts the CUCB-SW algorithm for scheduling while learning the non-stationary mean service rate. Thus, the regret $\Rr(\tau_j)$ as in \eqref{original_regret1} serves to capture the learning efficiency of CUCB-SW with theoretical guarantee in \cite{combMAB_nonstationary1}. To this end, the regret bound under our choice of parameter $d= \tau^{\frac{2}{3}(1-\alpha)}$ is given by Lemma \ref{lma:regret_bound} (in Appendix \ref{append_drift_bound1}) as follows:
\begin{align}
    \Rr(\tau_j)\leq c_0 \log(\tau) \tau^{\frac{1}{3} (\alpha + 2)}, \label{regret_bound_inter1}
\end{align}
for some universal constant $c_0 > 0$ that can be explicitly determined. Substituting \eqref{regret_bound_inter1} into \eqref{inter_drift2} and defining $c_1 = \frac{c_0}{\mu_{min}}$, we have:
    \begin{align}
\nonumber
    \Delta (\tau_j) &\leq B_1 \tau (\tau+1) +  2 (\mu_{max} + \varepsilon \mu_{min})\tau \EE[X_{\tau_{j}}]   \\
    &\quad - 2   \tau \| \QQ(\tau_j) \|_\infty \mu_{min} \big( \varepsilon  -  c_1 \log(\tau) \tau^{\frac{1}{3} (\alpha -1)} \big). \label{inter_drift3}
\end{align}
Now, if $ \varepsilon \geq c_1 \log(\tau) \tau^{\frac{1}{3}(\alpha -1) } $, we obtain from \eqref{inter_drift3} that:
\begin{align*}
    \Delta (\tau_j) &\leq B_1 \tau (\tau+1) +  2 (\mu_{max} + \varepsilon \mu_{min})\tau \EE[X_{\tau_{j}}]. 
\end{align*}
Recall from \eqref{lyapunov_drift1} that $ \Delta (\tau_j) = \EE\big[\Phi(\WQQ(\tau_{j+1})) - \Phi(\WQQ(\tau_{j})) \big|  \WQQ(\tau_{j}), \QQ(\tau_{j}) \big]$. Taking expectation on both sides of the above with respect to $ \WQQ(\tau_{j})$ and $ \QQ(\tau_{j})$, we get:
\begin{align*}
    &\EE\big[\Phi(\WQQ(\tau_{j+1})) \big] - \EE\big[\Phi(\WQQ(\tau_{j})) \big] \\
    &\leq  B_1 \tau (\tau+1)  +  2 (\mu_{max} + \varepsilon \mu_{min})\tau \EE[X_{\tau_{j}}]. 
\end{align*}
Summing up the above for $j = 0, 1, ..., K-1$ and noting that $ \Phi(\WQQ(\tau_{0})) = \Phi(\WQQ(0)) = 0$, we obtain that: 
\begin{align}
\nonumber
  \EE\big[\Phi(\WQQ(\tau_{K})) \big]  &\leq     B_1 \tau (\tau+1) K +   2 (\mu_{max} + \varepsilon \mu_{min})\tau \sum_{j=0}^{K-1} \EE[X_{\tau_{j}}] \\
  \nonumber
  &\overset{\eqref{shedding_cost}}{=} O\big( \tau^2 K + \tau \sum_{j=0}^{K-1} ( \tau_{j}^{\beta+1} \tau^{-1} + \varepsilon \tau_{j} ) \big)  \\
  \nonumber
  &=  O\big( \tau^2 K + \tau^{\beta+1} \sum_{j=0}^{K-1}    j^{\beta+1}  + \varepsilon \tau^2  \sum_{j=0}^{K-1}j  \big) \\
  \nonumber
  &=O\big( \tau^2 K + \tau^{\beta+1} \int_{0}^{K-1}    x^{\beta+1}  dx + \varepsilon \tau^2  \sum_{j=0}^{K-1}j  \big)  \\
  \nonumber
  &=O\big( \tau^2 K + \tau^{\beta+1}K^{\beta+2}   + \varepsilon \tau^2  K^2 \big)\\
  &=O\big( \tau T + T^{\beta+2}  \tau^{-1}   + \varepsilon T^2 \big), \label{potential_bound1}
\end{align}
where for last line, we recall that $K$ is the largest number such that $\tau_K = \tau K < T$. From Lemma \ref{lma:queue_to_potential}  (in Appendix \ref{append_drift_bound1}), we have:
\begin{align}
\nonumber
     \EE\big[\sum_{e\in E} \tq_e(\tau_{K}) \big] &\leq \sqrt{|E| \cdot \EE\big[ \Phi(\WQQ(\tau_{K})) \big]}\\
     &\overset{\eqref{potential_bound1}}{=} O\big( T^{\frac{1}{2}}\tau^{\frac{1}{2}} + T^{1+\frac{\beta}{2}}  \tau^{-\frac{1}{2}}  + \varepsilon^{\frac{1}{2}} T \big). \label{potential_bound2}
\end{align}
Noting that $0< T -\tau_K \leq \tau$ and using Lemma \ref{lma:general_queue_properties} (in Appendix \ref{append_drift_bound1}), we have:
\begin{align*}
     \EE\big[\sum_{e\in E} \tq_e(T) \big] &=  \EE\big[\sum_{e\in E} \tq_e(\tau_{K}) \big]  +  \EE\big[\sum_{e\in E} ( \tq_e(T) - \tq_e(\tau_{K}) ) \big] \\
     &\leq  \EE\big[\sum_{e\in E} \tq_e(\tau_{K}) \big] + (T - \tau_K)  (A_{max} + \mu_{max}) \\
     &\overset{\eqref{potential_bound2}}{=} O\big( T^{\frac{1}{2}}\tau^{\frac{1}{2}} + T^{1+\frac{\beta}{2}}  \tau^{-\frac{1}{2}}  + \varepsilon^{\frac{1}{2}} T  + \tau \big)\\
     &=O\big( T^{\frac{1}{2}}\tau^{\frac{1}{2}} + T^{1+\frac{\beta}{2}}  \tau^{-\frac{1}{2}}  + \varepsilon^{\frac{1}{2}} T \big),
\end{align*}
which concludes the proof of the Lemma.

\subsection{Supplementary Lemmas}
\label{append_drift_bound1}

\begin{lemma}
\label{lma:general_queue_properties}
    We have the following bounds $\forall e \in E$:
\begin{align}
    |Q_e(t_1) - Q_e(t_2)| \leq |t_1 - t_2| (A_{max} + \mu_{max}) \label{relate_vq1}\\
    |\tq_e(t_1) - \tq_e(t_2)| \leq |t_1 - t_2| (A_{max} + \mu_{max}) \label{relate_vq2}\\
    \tq_e(t) \leq Q_e(t) \leq \tq_e(t) + \sum_{q= 0}^{t-1} (a_e(q) - \tilde{a}_e(q)) \label{relate_vq3}  \\
    \|\WQQ(t)\|_\infty \leq \|\QQ(t)\|_\infty \leq   \|\WQQ(t)\|_\infty + X_t \label{relate_vq4}
\end{align}

\end{lemma}
\begin{proof}
\eqref{relate_vq1} trivially holds for $t_1 = t_2$. If $t_1 \neq t_2$ , WLOG, we assume that $t_1 > t_2$. From the queue dynamics \eqref{original_sys_queue}, we have:
\begin{align*}
    Q_e(t_1) &\geq Q_e(t_1-1) + a_e(t_1-1) - b_e(t_1-1)  \\
    &\geq Q_e(t_1-1) -\mu_{max}, \\
    Q_e(t_1) &\leq Q_e(t_1-1) + a_e(t_1-1)\leq Q_e(t_1-1) + A_{max}, 
\end{align*}
where we use $0\leq a_e(t) \leq A_{max}$ and $ 0\leq b_e(t) \leq \theta_e(t) \leq \mu_{max}$.
Iterating the above, we obtain that:
\begin{align*}
    &{Q}_e(t_1) \geq {Q}_e(t_2) - (t_1 - t_2) \mu_{max},\\
    &{Q}_e(t_1) \leq {Q}_e(t_2) + (t_1 - t_2) A_{max}.
\end{align*}
Combining the two above, we have \eqref{relate_vq1}. Similarly, we obtain \eqref{relate_vq2}.   

Next we proceed to prove \eqref{relate_vq3} by induction om $t$.

\textit{Base case $t= 0$:} Now, \eqref{relate_vq3} trivially holds since $ \tq_e(0) = Q_e(0) = 0, \forall e\in E$.

\textit{Inductive step $t \to t+1$:} First, we note that the number of packet arrivals of the imaginary system are shed from and thus upper-bounded by the number of packet arrivals   of the original system, i.e. $ \tilde{a}_e(q) \leq a_e(q), \forall e\in E, q= 0, 1, 2,...$
From the inductive hypothesis $ \tq_e(t) \leq Q_e(t) \leq \tq_e(t) + \sum_{q= 0}^{t-1} (a_e(q) - \tilde{a}_e(q))$ and the queue dynamics \eqref{original_sys_queue} and \eqref{imaginary_sys_queue}, we have $\forall e \in E$:
\begin{align*}
 Q_e(t+1) &=  \big(Q_e(t) + a_e(t) - b_e(t) \big)^+\\
 &\geq \big(\tq_e(t) + a_e(t) - b_e(t) \big)^+ \\
 &\overset{\eqref{imaginary_sys_queue}}{=} \tq_e(t+1),
\end{align*}
and,
\begin{align*}
    Q_e(t+1) &=  \big(Q_e(t) + a_e(t) - b_e(t) \big)^+\\
 &\leq \big(\tq_e(t) + \sum_{q= 0}^{t-1} (a_e(q) - \tilde{a}_e(q)) + a_e(t) - b_e(t) \big)^+ \\
 &\leq \big(  \tq_e(t) + \tilde{a}_e(t) - b_e(t)  + \sum_{q= 0}^{t} (a_e(q) - \tilde{a}_e(q)) \big)^+ \\
 &\leq \big(  \tq_e(t) + \tilde{a}_e(t) - b_e(t)   \big)^+ + \sum_{q= 0}^{t} (a_e(q) - \tilde{a}_e(q))\\
 &\overset{\eqref{imaginary_sys_queue}}{=} \tq_e(t+1) + \sum_{q= 0}^{t} (a_e(q) - \tilde{a}_e(q)).
\end{align*}
Thus, \eqref{relate_vq3} also holds for $t+1$. 

To prove \eqref{relate_vq4}, we first obtain from \eqref{relate_vq3} that:
\begin{align*}
    \tq_e(t) \leq Q_e(t) \leq \tq_e(t) + X_t, \forall e\in E.
\end{align*}
Next, we consider $e_1^* = \argmax_{e\in E}\tq_e(t)$ and $e_2^* = \argmax_{e\in E} Q_e(t)$ and get that:
\begin{align*}
    &\| \WQQ(t)\|_\infty =  \tq_{e_1^*}(t) \leq  Q_{e_1^*}(t) \leq  \| \QQ(t)\|_\infty,\\
    &\| \QQ(t)\|_\infty =  Q_{e_2^*}(t) \leq  \tq_{e_2^*}(t) + X_t \leq \| \WQQ(t)\|_\infty + X_t,
\end{align*}
which concludes the proof of the Lemma.
\end{proof}

\begin{lemma}
\label{lma:drift_bound1}
We have the following bound:
\begin{align}
\nonumber
     \Delta (\tau_j) &\leq B_1 \tau (\tau+1) + 2 \sum_{t= \tau_j}^{\tau_{j+1}-1} \WQQ(\tau_j)^T \wlmb(t) \\
     &\quad  - 2 \sum_{t= \tau_j}^{\tau_{j+1}-1} \EE[ \WQQ(\tau_j)^T \bB(t)\big| \WQQ(\tau_{j}), \QQ(\tau_{j}) ], \label{drfit1}
\end{align}
where $B_1 = \frac{1}{2} |E|  (A_{max} + \mu_{max})^2$.
\end{lemma}

\begin{proof}
From the queue process \eqref{imaginary_sys_queue} of the imaginary system, we first obtain that $\forall e \in E, t\in [\tau_j, \tau_{j+1})$:
\begin{align*}
   \tq_e(t+1)^2 &\leq   \big(\tq_e(t) + \tilde{a}_e(t) - b_e(t) \big)^2 \\
   &= \tq_e(t)^2  + ( \tilde{a}_e(t) - b_e(t))^2 + 2 \tq_e(t) ( \tilde{a}_e(t) - b_e(t)) \\
   &= \tq_e(t)^2  + ( \tilde{a}_e(t) - b_e(t))^2 + 2 \tq_e(\tau_j) ( \tilde{a}_e(t) - b_e(t)) \\
   &\quad  + 2 (\tq_e(t) - \tq_e(\tau_j)) ( \tilde{a}_e(t) - b_e(t))\\ 
   &\leq \tq_e(t)^2  + ( \tilde{a}_e(t) - b_e(t))^2 + 2 \tq_e(\tau_j) ( \tilde{a}_e(t) - b_e(t)) \\
   &\quad  + 2 |\tq_e(t) - \tq_e(\tau_j)| | \tilde{a}_e(t) - b_e(t)|\\ 
   &\leq \tq_e(t)^2 + (A_{max} + \mu_{max})^2 (1 + t - \tau_j) \\
   &\quad + 2 \tq_e(\tau_j) ( \tilde{a}_e(t) - b_e(t)),
\end{align*}
where in the last line we use  $0\leq \tilde{a}_e(t) \leq a_e(t) \leq A_{max}$, $ 0\leq b_e(t) \leq \theta_e(t) \leq \mu_{max}$ and Lemma \ref{lma:general_queue_properties} which gives $ |\tq_e(t) - \tq_e(\tau_j)| \leq (t- \tau_j) (A_{max} + \mu_{max})$. Telescoping the above for $t = \tau_j \to \tau_{j+1}-1$ and summing over all $e\in E$, we obtain that: 
\begin{align*}
    \sum_{e\in E} \tq_e(\tau_{j+1})^2 &\leq \sum_{e\in E}  \tq_e(\tau_j)^2  + \frac{1}{2} |E|  (A_{max} + \mu_{max})^2 \tau (\tau+1) \\
    &\quad 2 \sum_{e\in E} \tq_e(\tau_j) ( \tilde{a}_e(t) - b_e(t)).
\end{align*}
Taking the expectation conditioned on  $\WQQ(\tau_j)$ and $\QQ(\tau_j)$  of the above and noting that $\tilde{\lambda}_e(t) =\EE[ \tilde{a}_e(t)] = \EE[ \tilde{a}_e(t) \big|\WQQ(\tau_j), \QQ(\tau_j) ]$, i.e. the packet arrivals are independent of the queue lengths, we conclude the proof of \eqref{drfit1}.
\end{proof}

\begin{lemma}
\label{lma:relate_imaginary_regret1}
 We have the following bound:
 \begin{align}
     \nonumber 
  &\sum_{t= \tau_j}^{\tau_{j+1}-1} \EE\big[ \WQQ(\tau_j)^T \bB(t)  \big| \WQQ(\tau_{j}), \QQ(\tau_{j}) \big] \\
  \nonumber 
  &\geq  (1-\varepsilon) \sum_{t= \tau_j}^{\tau_{j+1}-1} \EE\big[ \WQQ(\tau_j)^T \bB^{\pi_0}(t)  \big| \WQQ(\tau_{j}), \QQ(\tau_{j}) \big] \\
  &\quad  + \varepsilon  \tau \| \WQQ(\tau_j) \|_\infty \mu_{min} - \| \WQQ(\tau_j) \|_\infty   \bar{\Rr}(\tau_j) \label{relate_imaginary_regret1}
 \end{align}
\end{lemma}

\begin{proof}
First, letting $e^* = \argmax_{e\in E} \tq_e(\tau_j)$,we note that:
\begin{align}
\nonumber
      &\sum_{t= \tau_j}^{\tau_{j+1}-1} \max_{\xX'\in \Mm} \EE\big[\sum_{e\in E}  \tq_e(\tau_j) x'_e \theta_e(t)  \big|\WQQ(\tau_{j}), \QQ(\tau_{j}) \big] \\
      \nonumber 
      &\geq \sum_{t= \tau_j}^{\tau_{j+1}-1} \| \WQQ(\tau_j) \|_\infty \cdot 1 \cdot  \EE\big[ \theta_{e^*}(t)  \big|\WQQ(\tau_{j}), \QQ(\tau_{j}) \big]]\\
      &\geq \sum_{t= \tau_j}^{\tau_{j+1}-1} \| \WQQ(\tau_j) \|_\infty \mu_{min} = \tau \| \WQQ(\tau_j) \|_\infty \mu_{min} \label{relate_imaginary_regret1_proof1}, 
\end{align}
where for the first inequality, we compare the maximizing solution with the feasible activation link vector that activates only the link $e^*$. Also, we have:
\begin{align}
\nonumber
      &\sum_{t= \tau_j}^{\tau_{j+1}-1} \max_{\xX'\in \Mm} \EE\big[\sum_{e\in E}  \tq_e(\tau_j) x'_e \theta_e(t)  \big|\WQQ(\tau_{j}), \QQ(\tau_{j}) \big] \\
      &\geq \sum_{t= \tau_j}^{\tau_{j+1}-1} \EE\big[ \WQQ(\tau_j)^T \bB^{\pi_0}(t)  \big| \WQQ(\tau_{j}), \QQ(\tau_{j}) \big] \label{relate_imaginary_regret1_proof2}, 
\end{align}
where we compare the maximizing policy with the policy $\pi_0$ with also the full knowledge of  every link $e$'s weight, i.e. 
$$ \EE\big[ \tq_e(\tau_j) \theta_e(t) \big| \WQQ(\tau_{j}), \QQ(\tau_{j}) \big] =  \tq_e(\tau_j) \mu_e(t).$$
Now, from \eqref{imaginary_regret2}, we have:     
\begin{align*}
   &\sum_{t= \tau_j}^{\tau_{j+1}-1} \EE\big[ \WQQ(\tau_j)^T \bB(t)  \big| \WQQ(\tau_{j}), \QQ(\tau_{j}) \big] \\
   &=   \sum_{t= \tau_j}^{\tau_{j+1}-1} \max_{\xX'\in \Mm} \EE\big[\sum_{e\in E}  \tq_e(\tau_j) x'_e \theta_e(t)  \big|\WQQ(\tau_{j}), \QQ(\tau_{j}) \big]  -\| \WQQ(\tau_j) \|_\infty   \bar{\Rr}(\tau_j)  \\
   &\overset{\eqref{relate_imaginary_regret1_proof1} + \eqref{relate_imaginary_regret1_proof2} }{\geq} (1-\varepsilon) \sum_{t= \tau_j}^{\tau_{j+1}-1} \EE\big[ \WQQ(\tau_j)^T \bB^{\pi_0}(t)  \big| \WQQ(\tau_{j}), \QQ(\tau_{j}) \big] \\
  &\quad  + \varepsilon  \tau \| \WQQ(\tau_j) \|_\infty \mu_{min} - \| \WQQ(\tau_j) \|_\infty   \bar{\Rr}(\tau_j),
\end{align*}
which concludes the proof of the Lemma.
\end{proof}

\begin{lemma}
\label{lma:relate_two_regrets}
    We have the following bound:
    \begin{align}
    \| \WQQ(\tau_j) \|_\infty   \bar{\Rr}(\tau_j) \leq \| \QQ(\tau_j) \|_\infty   \Rr(\tau_j) + \tau \mu_{max} \EE[X_{\tau_j}] \label{relate_two_regrets}
\end{align}
\end{lemma}
\begin{proof}
 From \eqref{imaginary_regret2}, we have:   
 \begin{align}
 \nonumber 
      \| \WQQ(\tau_j) \|_\infty   \bar{\Rr}(\tau_j) &=    \sum_{t= \tau_j}^{\tau_{j+1}-1} \max_{\xX'\in \Mm} \EE\big[\sum_{e\in E}  \tq_e(\tau_j) x'_e \theta_e(t)  \big|\WQQ(\tau_{j}), \QQ(\tau_{j}) \big]  \\
     &\quad -\sum_{t= \tau_j}^{\tau_{j+1}-1} \EE\big[ \WQQ(\tau_j)^T \bB(t)  \big| \WQQ(\tau_{j}), \QQ(\tau_{j}) \big] .\label{relate_two_regrets_proof1}
 \end{align}
 From Lemma \ref{lma:general_queue_properties}, we have $\forall e\in E$:
 \begin{align}
 \nonumber 
     &\EE\big[Q_e(\tau_j) \big| \WQQ(\tau_{j}), \QQ(\tau_{j}) \big]  - \sum_{t=0}^{\tau_j-1} (\lambda_e(t) - \tilde{\lambda}_e(t)  ) \\
     &\leq \EE\big[\tq_e(\tau_j) \big| \WQQ(\tau_{j}), \QQ(\tau_{j}) \big]  \leq \EE\big[Q_e(\tau_j) \big| \WQQ(\tau_{j}), \QQ(\tau_{j}) \big].\label{relate_two_regrets_proof2}
 \end{align}
 Plugging \eqref{relate_two_regrets_proof2} into \eqref{relate_two_regrets_proof1}, we have:
 \begin{align}
 \nonumber 
      \| \WQQ(\tau_j) \|_\infty   \bar{\Rr}(\tau_j) &=    \sum_{t= \tau_j}^{\tau_{j+1}-1} \max_{\xX'\in \Mm} \EE\big[\sum_{e\in E}  \tq_e(\tau_j) x'_e \theta_e(t)  \big|\WQQ(\tau_{j}), \QQ(\tau_{j}) \big]  \\
      \nonumber
     &\quad -\sum_{t= \tau_j}^{\tau_{j+1}-1} \EE\big[ \WQQ(\tau_j)^T \bB(t)  \big| \WQQ(\tau_{j}), \QQ(\tau_{j}) \big] \\
\nonumber 
&\leq     \sum_{t= \tau_j}^{\tau_{j+1}-1} \max_{\xX'\in \Mm} \EE\big[\sum_{e\in E}  Q_e(\tau_j) x'_e \theta_e(t)  \big|\WQQ(\tau_{j}), \QQ(\tau_{j}) \big]  \\
\nonumber
     &\quad -\sum_{t= \tau_j}^{\tau_{j+1}-1} \EE\big[ \QQ(\tau_j)^T \bB(t)  \big| \WQQ(\tau_{j}), \QQ(\tau_{j}) \big] \\
    \nonumber
    & + \sum_{t= \tau_j}^{\tau_{j+1}-1} \sum_{t=0}^{\tau_j-1}  \sum_{e\in E} (\lambda_e(t) - \tilde{\lambda}_e(t)  ) \EE\big[b_e(t) \big| \WQQ(\tau_{j}), \QQ(\tau_{j}) \big]  \\
    \nonumber 
    &\overset{\eqref{original_regret2}}{\leq}  \| \QQ(\tau_j) \|_\infty   \Rr(\tau_j)  + \tau \mu_{max} \EE[X_{\tau_j}],
 \end{align}
 which concludes the proof of the Lemma. 
\end{proof}

\noindent 
\textbf{Lemma \ref{lma:regret_bound}. [Restated]} Under MW-UCB,  the regret $\Rr(\tau_j)$ can be bounded by:
\begin{align*}
    \Rr(\tau_j) &\leq |E|\bigg( \frac{\tau}{d} +1\bigg) \bigg(2\sqrt{6 \log(\tau)} + 48 \sqrt{d} \log(\tau)  \bigg) \\
    &\quad +4 |E|  d \cdot \gamma(\tau_j, \tau_{j+1})  + |E|   \frac{\tau}{\sqrt{d}}+ \frac{\pi^2}{3} |E|^2 \mu_{max} \\
    &\quad+ \frac{\pi^2}{6} |E|^2 \mu_{max}    \log\big(  2 d^{1/2}  \big). 
\end{align*}
Under  Assumption \ref{vari_assumption} and by setting  $d= \Theta(\tau^{\frac{2}{3}(1-\alpha)})$, we further have  $\Rr(\tau_j) = O\big( \log(\tau) \tau^{\frac{1}{3} (\alpha + 2)}  \big)$.
\begin{proof}
When we fix the queue lengths $\QQ(\tau_j)$ throughout the frame $[\tau_j, \tau_{j+1})$, we aim to find a scheduling policy that solves 
\begin{align*}
&\max_{\xX'\in \Mm} \EE\big[\sum_{e\in E} w_e(\tau_j) x'_e \theta_e(t)  \big|\WQQ(\tau_{j}), \QQ(\tau_{j}) \big] \\
= &\max_{\xX'\in \Mm} \sum_{e\in E}x'_e  \mu_e(t) \EE\big[ w_e(\tau_j) \big|\WQQ(\tau_{j}), \QQ(\tau_{j}) \big], 
\end{align*}
over the $\tau$ time slots from $\tau_j$ to $\tau_{j+1}$ despite not knowing $\muu(t)$ and thus the true rewards at the time of making decisions. This problem, whereby the mean reward of each arm, i.e. $\mu_e(t) w_e(\tau_j)$, varies over time, can be characterized as stochastic combinatorial multi-armed bandit (SCMAB) problem in non-stationary environment and solved via the CUCB-SW algorithm \cite{combMAB_nonstationary1}. To derive the bound for $\Rr(\tau_j)$, we first verify the conditions required by \cite{combMAB_nonstationary1} and adapt the notations therein to our case. In particular, our model corresponds to SCMAB without probabilistically triggered arms, each of which is associated with a link $e \in E$. At any time slot $t \in [\tau_j, \tau_{j+1})$, an action is a link activation vector $\xX \in \Mm$. The expected reward of arm $e\in E$ at time $t \in [\tau_j, \tau_{j+1})$, if it's activated, is denoted by $W_e(t) = w_e(\tau_j) \mu_e(t)$. Let $\WW(t) = ( W_e(t))_{e\in E}$ be the vector of the arms' expected rewards at time $t$. The total variation of the mean reward statistics inside the frame $[\tau_j, \tau_{j+1})$ is thus depicted by:
\begin{align}
    V(\tau_j) &=   \sum_{t= \tau_j + 1}^{\tau_{j+1}} \|\WW(t) - \WW(t-1) \|_\infty \label{variation1}\\
    \nonumber 
    &= \sum_{t= \tau_j + 1}^{\tau_{j+1}} \| \big( w_e(\tau_j) (\mu_e(t) - \mu_e(t-1) \big) )_{e\in E}\|_\infty\\
    &\leq \sum_{t= \tau_j + 1}^{\tau_{j+1}} \|\muu(t) - \muu(t-1) \|_\infty  \label{variation2}  \\
    &= \gamma(\tau_j, \tau_{j+1})  = O(\tau^\alpha) \label{variation3},
\end{align}
where \eqref{variation2} holds since $w_e(\tau_j) \leq 1$, and \eqref{variation3} is by Assumption \ref{vari_assumption}. For convenience, we denote the total expected reward under the  arms' expected rewards $\WW$ and the action $\xX \in \Mm$ as:
\begin{align}
    r(\WW, \xX) = \sum_{e\in E} W_e x_e.
\end{align}
In view of the requirements imposed by \cite{combMAB_nonstationary1}, given two vectors of expected rewards $\WW$ and $ \WW'$ and any action $\xX$, we can verify that our model satisfies both the $\ell_1$ TPM bounded smoothness assumption of  with constant $B=1$, i.e. 
\begin{align*}
    |r(\WW, \xX) - r(\WW', \xX)| \leq \sum_{e: x_e = 1} |W_e - W_e' |,
\end{align*}
and the monotonicity assumption, i.e. if $\WW \leq \WW'$ (entry-wise), we have:
\begin{align*}
    r(\WW, \xX) \leq r(\WW', \xX).
\end{align*}
Furthermore, for each action $\xX \in \Mm$, the optimality gap with respect to the reward $\WW$ is defined as $\Delta_\xX^\WW = \max_{\xX' \in \Mm} r(\WW, \xX') - r(\WW, \xX)$. Then for each arm $e\in E$ and $t\in [\tau_j, \tau_{j+1})$, we define:
\begin{align*}
    \Delta^{e,t}_{min} = \min_{\xX \in \Mm: \Delta_\xX^{\WW(t)} > 0 } \Delta_\xX^{\WW(t)}, \\
    \Delta^{e,t}_{max} = \max_{\xX \in \Mm: \Delta_\xX^{\WW(t)} > 0 } \Delta_\xX^{\WW(t)}.
\end{align*}
We define $ \Delta^{e,t}_{min} = \infty$ and $ \Delta^{e,t}_{max} = 0$ if they are not properly defined by the above definitions. Then, $\Delta_{min} = \inf_{e\in E, t\in [\tau_j, \tau_{j+1})}  \Delta^{e,t}_{min}$ and $\Delta_{max} = \sup_{e\in E, t\in [\tau_j, \tau_{j+1})}  \Delta^{e,t}_{max}$ are respectively the minimum and maximum gap. For our problem instance, noting that $W_e(t) = w_e(\tau_j) \mu_e(t)\leq \mu_{max}$, we have the following bounds on the optimality gaps:
\begin{align}
\nonumber
     &\Delta_\xX^{\WW(t)} \leq  \max_{\xX' \in \Mm} r(\WW(t), \xX') \leq \sum_{e\in e} W_e(t) \leq |E| \mu_{max} \\
     \therefore &\Delta_{max} \leq |E| \mu_{max}. \label{delta_max_bound}
\end{align}
Following \cite{combMAB_nonstationary1}, given a set of positive parameters $\{M_e\}_{e\in E}$ and for any action $\xX\in \Mm$, we define $M_{\xX} = \max_{e: x_e = 1} M_e$ with the convention that $M_{\xX} = 0$ if $x_e = 0, \forall e\in E$. From the proof of \cite[Theorem 4]{combMAB_nonstationary1} , we have the following regret bound given the frame size $\tau$, sliding-window size $d$ and any arbitrary set of positive parameters $\{M_e\}_{e\in E}$: 
\begin{align}
\nonumber
    \Rr(\tau_j) &\leq \sum_{e\in E} \bigg(\frac{\tau}{d}+1\bigg) \bigg(2\sqrt{6 \log(\tau)} + \frac{48|E| \log(\tau)}{M_e} \bigg) \\
\nonumber
    &\quad +4 |E| d \cdot V(\tau_j)  + \sum_{t=\tau_j}^{\tau_{j+1}-1} M_{\xX(t)} + \frac{\pi^2}{3} |E|  \Delta_{max} \\
    &\quad+ \frac{\pi^2}{6}  \Delta_{max} \sum_{e\in E} j^e_{max}, \label{regret_bound_step1}
\end{align}
where $j^e_{max} = \max\big\{ \lceil \log\big(  \tfrac{2 |E|}{M_e}   \big) \rceil, 0 \big\} $ (also see \cite{combMAB_nonstationary4}). 
Finally, by plugging \eqref{variation3} and \eqref{delta_max_bound} into \eqref{regret_bound_step1} and setting  $M_e = M = \frac{|E|}{\sqrt{d}} 
, \forall e \in E$, which also implies that $M_{\xX(t)} = M, \forall t\in [\tau_j, \tau_{j+1})$, we conclude the required statements of the Lemma.
\end{proof}

\begin{lemma}
\label{lma:queue_to_potential}
We have the following bound for any time slot $T$:
\begin{align}
    \EE\big[\sum_{e\in E} \tq_e(T) \big] \leq \sqrt{|E| \cdot \EE\big[ \Phi(\WQQ(T)) \big]}
\end{align}
\end{lemma}

\begin{proof}
 By Cauchy–Schwarz inequality, we first have:
 \begin{align*}
  \sum_{e\in E} \tq_e(T) \leq \sqrt{|E| \big( \sum_{e\in E} \tq_e(T)^2   \big)} = \sqrt{|E| \Phi(\WQQ(T)) }.
 \end{align*}
 Taking expectation of the above and by Jensen's inequality, we obtain that:
 \begin{align*}
      \EE\big[\sum_{e\in E} \tq_e(T) \big] &\leq \EE\big[ \sqrt{|E|  \Phi(\WQQ(T)) } \big] \leq \sqrt{|E| \cdot \EE\big[ \Phi(\WQQ(T)) \big]},
 \end{align*}
 which concludes the proof of the Lemma.
\end{proof}

\section{Proof of Corollary \ref{cor:strong_stability}}
\label{appen_strong_stability}

The proof follows as a side result of a special case of the proof of  Lemma \ref{lma:imag_queue_bound} (Appendix \ref{appen_imag_queue_bound}). In particular, since now $\{\lmb(t) \}_{t\geq 0} \in  \Lmb_s(\beta)$, we can consider the shedding scheme that sheds no traffic, i.e. $\tilde{a}_e(t) = a_e(t)$ for any $e\in E$ and $t\geq 0$ and thus the total amount of shed traffic from \eqref{XT} is:
\begin{align}
    \label{zero_XT}
    X_T = \sum_{t=0}^{T-1} \sum_{e\in E} a_e(t)-\sum_{t=0}^{T-1} \sum_{e\in E} \tilde{a}_e(t) = 0.
\end{align}
Then the original and imaginary systems are now the same where $\{\QQ(t)\}_{t\geq 0} \equiv \{\WQQ(t)\}_{t\geq 0}$  and the Lyapunov drift \eqref{lyapunov_drift1} can  be equivalently written as:
\begin{align}
\label{lyapunov_drift_corol}
    \Delta (\tau_j) = \EE\big[\Phi(\QQ(\tau_{j+1})) - \Phi(\QQ(\tau_{j})) \big|  \WQQ(\tau_{j}), \QQ(\tau_{j}) \big].
\end{align}
Now from \eqref{inter_drift3} in  the proof of  Lemma \ref{lma:imag_queue_bound}

 \begin{align}
\nonumber
    \Delta (\tau_j) &\leq B_1 \tau (\tau+1) +  2 (\mu_{max} + \varepsilon \mu_{min})\tau \EE[X_{\tau_{j}}]   \\
    \nonumber
    &\quad - 2   \tau \| \QQ(\tau_j) \|_\infty \mu_{min} \big( \varepsilon  -  c_1 \log(\tau) \tau^{\frac{1}{3} (\alpha -1)} \big) \\
    \nonumber
    &\overset{\eqref{zero_XT}}{=} B_1 \tau (\tau+1)  - 2   \tau \| \QQ(\tau_j) \|_\infty \mu_{min} \big( \varepsilon  -  c_1 \log(\tau) \tau^{\frac{1}{3} (\alpha -1)} \big)\\
    &\leq  B_1 \tau (\tau+1)  - 2   \tau \big( \sum_{e\in E} Q_e(\tau_j) \big)  |E|^{-1}  \mu_{min} \big( \varepsilon  -  c_1 \log(\tau) \tau^{\frac{1}{3} (\alpha -1)} \big), \label{strong_stability_proof1}
\end{align}
where for the last line we use   $ \sum_{e\in E} Q_e(\tau_j) \leq  \sum_{e\in E} \|\QQ(\tau_j)\|_\infty= |E| \|\QQ(\tau_j)\|_\infty$. Taking expectation on both sides of \eqref{strong_stability_proof1} with respect to $ \WQQ(\tau_{j})$ and $ \QQ(\tau_{j})$ in view of \eqref{lyapunov_drift_corol}, we get:
\begin{align}
\nonumber
    &\EE\big[\Phi(\QQ(\tau_{j+1})) \big] - \EE\big[\Phi(\QQ(\tau_{j})) \big] \\
    &\leq  B_1 \tau (\tau+1) - 2   \tau \EE\big[ \sum_{e\in E} Q_e(\tau_j) \big]  |E|^{-1}  \mu_{min} \big( \varepsilon  -  c_1 \log(\tau) \tau^{\frac{1}{3} (\alpha -1)} \big). \label{strong_stability_proof2}
\end{align}
Now, if $\varepsilon > c_1 \log(\tau) \tau^{\frac{1}{3} (\alpha -1)}$, by summing \eqref{strong_stability_proof2} from $j = 0 \to K-1$ and noting that  $\Phi(\QQ(\tau_K)) \geq 0$ and $\Phi(\QQ(0)) = 0$, we obtain that:
\begin{align}
    \sum_{j=0}^{K-1} \sum_{e\in E} \EE\big[  Q_e(\tau_j) \big] \leq \frac{B_1 (\tau+1)K |E| }{2 \mu_{min} \big(\varepsilon  -  c_1 \log(\tau) \tau^{\frac{1}{3} (\alpha -1)} \big)}. \label{strong_stability_proof3}
\end{align}
By Lemma \ref{lma:general_queue_properties} and noting that $\tau_K = \tau K < T$ and $T - \tau_K \leq \tau$, we have:
\begin{align*}
      \sum_{t=0}^{T-1}  \sum_{e\in E}  \EE\big[  Q_e(t) \big] &=  \sum_{j=0}^{K-1} \sum_{t=\tau_j}^{\tau_{j+1}-1}  \sum_{e\in E}  \EE\big[  Q_e(t) \big] \\
      &\quad +\sum_{t=\tau_K}^{T-1}  \sum_{e\in E}  \EE\big[  Q_e(t) \big] \\
      &\leq  \sum_{j=0}^{K-1} \sum_{t=\tau_j}^{\tau_{j+1}-1}  \bigg[ \sum_{e\in E}  \EE\big[  Q_e(\tau_j)\big] \\
      &\quad \quad \quad + (t-\tau_j) |E|(A_{max} + \mu_{max}) \bigg] \\
      &\quad +\sum_{t=\tau_K}^{T-1}  \bigg[ \sum_{e\in E}  \EE\big[  Q_e(\tau_{K-1}) \big]  \\
      &\quad \quad \quad + (t-\tau_{K-1}) |E|(A_{max} + \mu_{max}) \bigg] \\
      &\leq 2 \tau  \sum_{j=0}^{K-1} \sum_{e\in E} \EE\big[  Q_e(\tau_j) \big] \\
      &\quad + \frac{|E|(A_{max} + \mu_{max})}{2} \big( K\tau(\tau-1) + 2 \tau^2 -\tau     \big) \\
      &\overset{\eqref{strong_stability_proof3} }{\leq} \frac{B_1 (\tau+1)K |E| }{ \mu_{min} \big(\varepsilon  -  c_1 \log(\tau) \tau^{\frac{1}{3} (\alpha -1)} \big)} \\
      &\quad + \frac{|E|(A_{max} + \mu_{max})}{2} \big( K\tau(\tau-1) + 2 \tau^2 -\tau   \big) \\
      &\leq \frac{2 B_1 T |E| }{ \mu_{min} \big(\varepsilon  -  c_1 \log(\tau) \tau^{\frac{1}{3} (\alpha -1)} \big)} \\
      &\quad + \frac{|E|(A_{max} + \mu_{max})}{2} \big(T (\tau-1) + 2 \tau^2 -\tau   \big) \\
      \therefore  \frac{  \sum_{t=0}^{T-1}  \sum_{e\in E}  \EE\big[  Q_e(t) \big]}{T}&\leq \frac{2 B_1 |E| }{ \mu_{min} \big(\varepsilon  -  c_1 \log(\tau) \tau^{\frac{1}{3} (\alpha -1)} \big)}\\
      &\quad + \frac{|E|(A_{max} + \mu_{max})}{2} \big( \tau-1 + \frac{2 \tau^2 -\tau}{T}   \big).
\end{align*}
Taking $\limsup$ of the above, we conclude that required statement of the Corollary, i.e. for fixed $\tau$,
\begin{align*}
    \limsup_{T\to \infty} \frac{1}{T} \sum_{t=0}^{T-1} \sum_{e\in E}  \EE[ Q_e(t)] < \infty. 
\end{align*}


\section{Supplementaries for Simulations}
\label{appen_observation}

For the experimental setup in Section \ref{sec:exp}, we justify that $\EE[\gamma(t_1, t_2)] = O(|t_2 - t_1|^{1/2})$  for any $0\leq t_1 < t_2 \leq T$ for both settings of $\delta_t = \frac{0.5}{T^{1/2}}$ and $\delta_t = \frac{0.5}{(t+1)^{1/2}}$. Recall from \eqref{formula_variation1} that $  \gamma(t_1, t_2) = \sum_{t= t_1+1}^{t_2} \|\muu(t) - \muu(t-1) \|_\infty$.
Under our experimental setting, since $\mu_e(t), \mu_e(t-1)\in \{0.25, 0.75\}, \forall e\in E, t> 0$, we obtain that:
\begin{align*}
    \EE[ \|\muu(t) - \muu(t-1) \|_\infty ] &= 0.5 \cdot P(\text{$\mu_e(t)$ changes its state for some $e$})\\
    &= 0.5 \cdot \big[ 1 - (1-\delta_t)^{|E|}\big] \\
    &\leq 0.5 |E| \delta_t,
\end{align*}
where the last line follows Bernoulli's inequality. Summing up the above for $t = t_1+1 \to t_2$, we have:
\begin{align}
    \EE[\gamma(t_1, t_2)] \leq 0.5 |E| \sum_{t= t_1+1}^{t_2} \delta_t. \label{observation_proof1}
\end{align}
If  $\delta_t = \frac{0.5}{T^{1/2}}$, we obtain from \eqref{observation_proof1} that:
\begin{align*}
     \EE[\gamma(t_1, t_2)] \leq 0.25 |E| \frac{t_2 -t_1}{T^{1/2}} \leq 0.25 |E| \sqrt{t_2 - t_1} = O(|t_2 - t_1|^{1/2}).
\end{align*}
If  $\delta_t = \frac{0.5}{(t+1)^{1/2}}$, we obtain from \eqref{observation_proof1} that:
\begin{align*}
     \EE[\gamma(t_1, t_2)] &\leq 0.25 |E|  \sum_{t= t_1+1}^{t_2} \frac{1}{(t+1)^{1/2}} \\ 
     &\leq 0.25 |E| \int_{t= t_1+2}^{t_2+1} \frac{1}{x^{1/2}} dx \\
     &= 0.5 |E| (\sqrt{t_2+1} - \sqrt{t_1+2})\\ 
     &\leq  0.5 |E| \sqrt{t_2 - t_1} = O(|t_2 - t_1|^{1/2}). 
\end{align*}



\end{document}